\documentclass[prb,epsf,epsfig,showpacs,amsmath,amssymb,floatfix,superscriptaddress,a4paper]{article}
\usepackage{amsmath}
\usepackage[affil-it]{authblk}
\usepackage[toc,page]{appendix}
\usepackage{graphicx,pdflscape}
\usepackage{axodraw4j}
\usepackage{pstricks}
\usepackage{color}
\usepackage{hyperref}

\usepackage[toc,page]{appendix}
\usepackage{graphicx,pdflscape,epsf,epsfig,amsmath}
\usepackage{alltt,dsfont,amsmath,amssymb,bm}
\usepackage{float}

\newcommand{\vk}{\vec{k}}

\newcommand{\beq}{\begin{equation}}
\newcommand{\eeq}{\end{equation}}
\newcommand{\barray}{\begin{eqnarray}}
\newcommand{\earray}{\end{eqnarray}}
\newcommand{\lab}[1]{\label{#1}}
\newcommand{\disp}[1]{Eq.(\ref{#1})}

\newcommand{\nn}{\nonumber}

\begin{document}

\title{High temperature expansion for dynamical correlation functions in the infinite-U Hubbard Model}
\author{Edward Perepelitsky}
\affil{ Physics Department, University of California, Santa Cruz, CA 95064, USA}
\date{\today}

\maketitle

\begin{abstract}

We develop a diagrammatic approach for calculating the high temperature expansion of dynamic correlation functions, such as the electron Green's function and the time-dependent density-density and spin-spin correlation functions, for the infinite-U Hubbard Model with any number of spin species. The formalism relies on the use of restricted lattice sums, in which distinct vertices of the diagram represent distinct sites on the lattice. We derive a new formula for the restricted lattice sum of a disconnected diagram consisting of several connected components, and use it to prove the linked cluster theorem with respect to ``generalized connected diagrams", formed by overlapping the original connected components on the lattice. This enables us to express all quantities as a sum over these generalized connected diagrams. We compute the Green's function to $4^{th}$ order in $\beta t$ for the case of $m$ spin species on a d-dimensional hypercube by hand. We take the $m\to\infty$ limit, enabling us to obtain expressions for the Dyson-Mori self-energy to $4^{th}$ order in $\beta t$ for the case of an infinite number of spin species. This may have connections to slave boson techniques used for the study of this model. Our approach is computationally more efficient than any used previously for the calculation of the high temperature expansion of dynamic correlation functions, and high order results for both the Green's function and the time-dependent density-density and spin-spin correlation functions shall be presented in a separate paper \cite{Ehsannew}.

\end{abstract}

\thispagestyle{empty}

\section{Introduction\lab{sec1} }

\subsection{Motivation}
The Hubbard model described by the hamiltonian
\beq H = -\sum_{ij\sigma}t_{ij}c^\dagger_{i\sigma }c_{j\sigma} +U\sum_{i}n_{i\uparrow}n_{i\downarrow}- \mu\sum_{i}n_i, \eeq
is one of the simplest models of itinerant electrons interacting on a lattice. Its solution for large enough values of $\frac{U}{t}$ is believed to contain many of the exotic properties observed in real materials, such as the cuprates. Recently, there has been much interest in understanding the behavior of dynamic correlation functions in this model in the limit of extreme correlations, that is, when the onsite repulsive potential $U$ tends to infinity. In this case, the Hilbert space is Gutzwiller projected, and each site is restricted to single occupancy. This is referred to as the infinite-U Hubbard model, and its Hamiltonian can be written in terms of the Hubbard $X$ operators\cite{ECQL}.

\beq H = -\sum_{ij\sigma}t_{ij}X_i^{\sigma 0 }X_j^{0\sigma} - \mu\sum_{i\sigma}X_i^{\sigma\sigma}. \label{Hamiltonian}\eeq
The operator $X_i^{ab} = |a\rangle\langle b|$ takes the electron at  site $i$ from the state $|b\rangle$ to the state $|a\rangle$, where for a model with $m$ spin species, $|a\rangle$ and $|b\rangle$ are one of $m+1$ allowed states, $m$ single occupancy states corresponding to the $m$ spin species, and the zero occupancy state. For the usual case of $m=2$ these are the states $|\uparrow\rangle$, $|\downarrow\rangle$, or $|-\rangle$. Due to the non-canonical nature of the Hubbard $X$ operators, the limit of extreme correlations presents great conceptual and technical challenges.

The recently developed ``Extremely Correlated Fermi Liquid Theory" (ECFL)\cite{ECFL}\cite{Monster} due to Shastry, has made impressive progress towards understanding this limit. A key idea that has emerged from this theory is the particle-hole asymmetry in the spectral density of the electron Green's function, and the spectral density of the Dyson-Mori self-energy, which becomes more pronounced as the density $n\to1$\cite{ECFL,Anatomy,Asymm}. This breaks a previous paradigm, in which the Fermi-liquid state was always believed to be particle-hole symmetric. This asymmetry is fundamentally a consequence of the Gutzwiller projection in the extreme correlation limit. 

Recent DMFT studies of the Hubbard model have also emphasized the presence of this particle-hole asymmetry for large values of $U$\cite{Xengetal}. A recent comparative study between ECFL and DMFT has explored the connection between these two methods further, demonstrating that the asymmetry is captured naturally through the ECFL functional form, by assuming simple particle-hole symmetric behavior for the two auxiliary ECFL self-energies\cite{ECFLDMFT, larged}. A comparative study between the ECFL and Numerical Renormalization Group calculations for the infinite-U Anderson impurity model points to the same conclusion\cite{ECFLAM}. In terms of experimental implications, the particle-hole asymmetry  has important consequences for understanding the large Seebeck coefficient of strongly-correlated materials\cite{Xengetal}, and understanding the anomalous line shapes of Angle-Resolved Photoemission Spectroscopy experiments, as has been done through the ECFL functional form\cite{Gweon}\cite{Kazue}. Therefore, many different sources of theoretical and experimental impetus point towards the importance of the extreme correlation limit, and its theoretical and experimental implications.

\subsection{Previous work}
An important tool in the study of the Hamiltonian in \disp{Hamiltonian} is the high temperature expansion. In this expansion, various static and dynamic quantities such as the thermodynamic potential, the electron Green's function, and the (time-dependent) density-density and spin-spin correlation functions are expanded in the parameter $\beta t$, where $\beta$ is the inverse temperature. Some recent examples of the use of the high temperature expansion can be found in Refs. \cite{Moments} and \cite{DMFThighT}, in which it is used synergistically with ECFL and DMFT respectively. In Ref. \cite{Moments}, the high-temperature expansion for the electron Green's function in the infinite-$U$ Hubbard model is used to benchmark ECFL calculations. Furthermore, combined with insight gained from the asymmetric shape of the ECFL spectral function, it is used to study the evolution of the Fermi surface in the limit $n\to 1$. In \cite{DMFThighT}, the high temperature expansion and DMFT are used to study the thermodynamic properties of the Hubbard Model and its implications for cold atomic gases in optical lattices.

In the 1970's and 1980's, Plischke\cite{Plischke}, Kubo and Tada\cite{KuboTada} extended the methods of Betts et. al.\cite{Betts} (for the XY model) to the study the thermodynamic and ferromagnetic properties of the infinite-$U$ Hubbard model, through the calculation of the high temperature expansion of the thermodynamic potential and static correlation functions. Although series expansions usually seek to remove disconnected diagrams using the linked-cluster theorem \cite{Nozieres,Domb}, their approach contains disconnected diagrams and relies on the use of restricted lattice sums, in which distinct vertices of a diagram represent distinct sites on the lattice. In 1991, Metzner formulated the linked-cluster expansion around the atomic limit of the Hubbard model\cite{Metzner}, in which the lattice sums are unrestricted, and disconnected diagrams are explicitly eliminated from the formalism. Metzner's expansion applies to both static and dynamic quantities, such as the single particle Green's function, and higher order dynamic correlation functions. In spite of this, there have not been many numerical results for the high-temperature series for dynamic correlation functions. Some notable exceptions are presented in Refs. \cite{EhsanMetzner} and \cite{Pairault}, in which the Green's function is calculated to $8^{th}$ order for the infinite-$U$ Hubbard model, and $5^{th}$ order for the finite-U Hubbard model respectively, using the Metzner (or similar) formalism.

\subsection{Results}
In this work, we extend the method of Kubo et. al. to the calculation of {\bf dynamic correlation functions} for the infinite-$U$ Hubbard model. We introduce an improvement in the evaluation of the spin sum and signature of a diagram, which permits us to make this evaluation with greater ease and for an arbitrary number of spin species $m$. We also introduce a novel approach for dealing with the problem of disconnected diagrams, this is the main achievement of our work. Although this problem has been addressed in various ways in the context of high-temperature expansions of the Ising and Heisenberg models\cite{Domb}, and the methods adapted to the infinite-U Hubbard model by Kubo et. al., our approach is distinct from those taken previously. It has the advantage that it can be easily generalized to the case of dynamic correlation functions, as we do here. In our approach, first the connected diagrams are evaluated. Their contribution feeds into a temporal part and a spatial part, the latter consisting of the lattice sum of the diagram. Then, an arbitrary number of the connected diagrams are chosen to create a ``generalized connected diagram". The temporal contribution of this generalized connected diagram consists of the product of the temporal contributions of the constituent diagrams, and its spatial contribution consists of the lattice sum of all the ways that one can partially or fully overlap these constituent diagrams on the lattice. The linked cluster theorem is then proved to be valid with respect to the generalized connected components. Our method is computationally more efficient than any used previously, and we are therefore able to calculate the Green's function to fourth order in $\beta t$ by hand. Taking the $m \to \infty$ limit, we obtain expressions for the Dyson-Mori self-energy valid in the limit of infinite spin species, which may be interesting in the context of slave boson techniques \cite{ReadNewns,Coleman,KotliarLiu}. Numerical high order calculations for both the Green's function and the time-dependent density-density and spin-spin correlation functions shall be presented in a separate paper\cite{Ehsannew}.
\subsection{Outline of paper}
In sections \ref{diagramsz} and \ref{spinsign}, we develop diagrammatic rules for the partition function, and give examples of their use. In section \ref{linked-cluster} we discuss the linked-cluster theorem with respect to ``generalized connected components". In sections \ref{lsdisc} and \ref{thermopotential} we derive a formula for the restricted lattice sum of a disconnected diagram with $n$ of the original components and use this formula to prove the aforementioned linked-cluster theorem. We are thus able to write the thermodynamic potential as a sum of the contributions of the generalized connected components. In sections \ref{diagramsG} and \ref{calcGreen}, we extend the methods developed for the thermodynamic potential to derive diagrammatic rules for the Green's function. In particular, the linked-theorem is used to show that the partition function in the denominator of the Green's function cancels the disconnected diagrams consisting of several generalized connected components in the numerator. Hence, the Green's function is written as a sum of generalized connected components. In section \ref{Greenfour}, we give results for the Green's function to $4^{th}$ order in $\beta t$ for $m$ spin species on a d-dimensional hypercube. In section \ref{infinitespins}, we give the $4^{th}$ order results for the Dyson-Mori self-energy in the limit of infinite spin species i.e. $m\to\infty$. Finally, in section \ref{ddss}, we extend the formalism to the calculation of time-dependent  density-density and spin-spin correlation functions.

\section{ Expansion for the thermodynamic potential\lab{sec 2} }
\subsection{Diagrams for the partition function\lab{diagramsz}}
The partition function and thermodynamic potential are defined as
\barray Z = Tr(e^{-\beta \hat{H}});\;\;\;\;\;\;\Omega = -T\ln Z.  \label{z1}\earray
We write the Hamiltonian as
\beq \hat{H} = \hat{T}-\mu \hat{N}, \label{TN} \eeq
where $ \hat{T} $ is the kinetic energy operator, and $ \hat{N} $  is the number operator. Comparison with \disp{Hamiltonian} shows that
\barray
\hat{T} = -\sum_{ij\sigma}t_{ij}X_i^{\sigma 0 }X_j^{0\sigma};\;\;\;
\hat{N}= \sum_{i\sigma}X_i^{\sigma\sigma}.
\earray
The partition function (\disp{z1}) is then written as
\beq Z = Tr(e^{\beta\mu\hat{N}}e^{-\beta\hat{T}}), \label{z2} \eeq
where we have used the fact that the kinetic energy operator commutes with the number operator. Expanding $e^{-\beta\hat{T}}$, we obtain
\beq \frac{Z}{Z_0} = \sum\limits_{n=0}^\infty \frac{\beta^n}{n!} \sum\limits_{\substack{j_1j_1'\ldots j_nj_n'\\\sigma_1\ldots \sigma_n}} t_{j_1j_1'}\ldots t_{j_nj_n'}\langle X_{j_1'}^{\sigma_10}X_{j_1}^{0\sigma_1}\ldots X_{j_n'}^{\sigma_n0}X_{j_n}^{0\sigma_n}\rangle_0, \label{z3}\eeq
where we have used the definitions
\barray Z_0 \equiv Tr(e^{\beta\mu\hat{N}});\;\;\;
\langle \hat{O}\rangle_0 \equiv \frac{Tr(e^{\beta\mu\hat{N}}\hat{O})}{Z_0}. 
\earray
The creation and destruction operators in the expectation value in \disp{z3} will distribute amongst the various sites of the lattice with the following restrictions. For a given site, creation and destruction must alternate. There must be an even number acting on each site. In addition, $X_{j_p'}^{\sigma_p0}$ and $X_{j_p}^{0\sigma_p}$ operate on neighboring sites. Within the expectation value, the operators must be permuted from their current order so that all operators acting on a given site are next to each other. The sign of the diagram is determined by whether the necessary permutation is even or odd. Once this permutation is done, the expectation value factors into a product of single site expectation values for each of the sites being acted on. Suppose there are $p$ operators acting on a given site. Then, the expectation value for this site must have one of the 2 following forms:
\beq \langle X^{\sigma_10}X^{0\sigma_2}X^{\sigma_30}X^{0\sigma_4}\ldots X^{\sigma_{p-1}0}X^{0\sigma_p}\rangle_0 =\frac{\rho}{m}\delta_{\sigma_p\sigma_1}\delta_{\sigma_{p-1}\sigma_{p-2}}\ldots\delta_{\sigma_3\sigma_2},\label{fulldelta} \eeq
or
\beq \langle X^{0\sigma_1}X^{\sigma_20}X^{0\sigma_3}X^{\sigma_40}\ldots X^{0\sigma_{p-1}}X^{\sigma_p0}\rangle_0 =(1-\rho)\delta_{\sigma_p\sigma_{p-1}}\delta_{\sigma_{p-2}\sigma_{p-3}}\ldots\delta_{\sigma_2\sigma_1}, \label{emptydelta}\eeq
where $\rho \equiv \frac{me^{\beta\mu}}{1+me^{\beta\mu}}$.
This observation allows us to write down the rules for calculating the $n^{th}$ order contribution to $\frac{Z}{Z_0}$:
\begin{itemize}

\item[1]  Using lines labeled by $\sigma_i$ starting from $\sigma_n$ and going down to $\sigma_1$, draw all topologically distinct diagrams, such that each line emerges from one vertex and enters into another one. Each vertex can be one of 2 types, a filled circle $\bigotimes$ or an empty one $\bigcirc$. Every time a line is drawn, it can be attached to two existing vertices, or one may create new vertices for it to attach to. Multiple lines can go into the same vertex. However, the following rules must be satisfied at each step $i:n\rightarrow1 $ of the diagram process.

\subitem[a] 
For a filled circle with an odd number of lines attached to it, there must be one more coming out than going in. The opposite is true for an empty circle, one more going in than coming out.
\subitem[b]
For an empty or filled circle with an even number of lines, as many must go out as come in.
\subitem[c]
In the final diagram, all circles must have an even number of lines.

\item[2] Insert a factor of $\frac{\rho}{m}$ for each filled circle, and $1-\rho$ for each empty circle.

\item[3] Each vertex is a distinct site on the lattice with lines connecting nearest neighbors. Compute the multiplicity of each diagram over the entire lattice (with the restriction that vertices are distinct sites and their relative positions are as indicated by the lines of the graph). Insert a factor of $t^n$.

\item[4] Insert a factor of $\frac{\beta^n}{n!}$.

\item[5] Determine the spin sum and the sign of the diagram as follows.  At each site, pair the lines in the following way. For a full site with p lines,

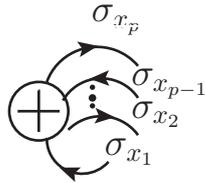
\begin{center}
\fcolorbox{white}{white}{
  \begin{picture}(213,86) (50,-30)
    \SetWidth{1.0}
    \SetColor{Black}
    \COval(75,27)(11,11)(90.0){Black}{White}\Line(67.222,27)(82.778,27)\Line(75,34.778)(75,19.222)

    \Arc[arrow,arrowpos=0.5,arrowlength=5,arrowwidth=2,arrowinset=0.2,clock](96.714,31.286)(19.882,160.263,39.753)
    \Text(95,56)[lb]{\Large{\Black{$\sigma_{x_p}$}}}
    \Arc[arrow,arrowpos=0.5,arrowlength=5,arrowwidth=2,arrowinset=0.2](98.308,22.89)(16.446,33.638,150.453)
    \Text(110,32)[lb]{\Large{\Black{$\sigma_{x_{p-1}}$}}}
    \Arc[arrow,arrowpos=0.5,arrowlength=5,arrowwidth=2,arrowinset=0.2,clock](98.158,8.057)(17.043,135.512,35.691)
    \Text(110,21)[lb]{\Large{\Black{$\sigma_{x_2}$}}}
    \Arc[arrow,arrowpos=0.5,arrowlength=5,arrowwidth=2,arrowinset=0.2,clock](91.106,16.603)(13.112,-41.007,-181.734)
    \Text(100,7)[lb]{\Large{\Black{$\sigma_{x_1}$}}}
    \Vertex(95,32){1.414}
    \Vertex(95,36){1}
    \Vertex(95,29){1}
    \Text(178,37)[lb]{\Large{\Black{$x_1<x_2\ldots <x_{p-1}<x_p$}}}
  \end{picture}
}
\end{center}

the pairings are $(x_1,x_p)(x_2,x_3)\ldots(x_{p-2},x_{p-1})$. For an empty site with p lines,

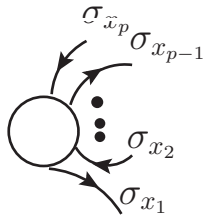
\begin{center}
\fcolorbox{white}{white}{
  \begin{picture}(160,89) (196,-80)
    \SetWidth{1.0}
    \SetColor{Black}
    \COval(210,-30)(13,13)(0){Black}{White}
    \Arc[arrow,arrowpos=0.5,arrowlength=5,arrowwidth=2,arrowinset=0.2,clock](233.667,-29.333)(13.437,-46.005,-150.255)
    \Arc[arrow,arrowpos=0.5,arrowlength=5,arrowwidth=2,arrowinset=0.2,clock](208.645,-79.269)(35.428,84.567,31.042)
    \Arc[arrow,arrowpos=0.5,arrowlength=5,arrowwidth=2,arrowinset=0.2](237.938,-17.984)(25.016,118.502,175.45)
    \Arc[arrow,arrowpos=0.5,arrowlength=5,arrowwidth=2,arrowinset=0.2,clock](240.343,-26.06)(21.227,163.413,82.81)
    \Vertex(230,-19){2.236}
    \Vertex(231,-27){2}
    \Vertex(231,-32){2}
    \Text(225,6)[lb]{\Large{\Black{$\sigma_{x_p}$}}}
    \Text(243,-4)[lb]{\Large{\Black{$\sigma_{x_{p-1}}$}}}
    \Text(242,-40)[lb]{\Large{\Black{$\sigma_{x_2}$}}}
    \Text(239,-61)[lb]{\Large{\Black{$\sigma_{x_1}$}}}
    \Text(321,-18)[lb]{\Large{\Black{$x_1<x_2<\ldots<x_{p-1}<x_p$}}}
  \end{picture}
}
\end{center}

the pairings are $(x_1,x_2)(x_3,x_4)\ldots(x_{p-1},x_{p})$. 

Split the diagram into loops in the following way. Find the filled site with the line labeled by $\sigma_n$ coming out of it. Find the line that this line is paired to on this site, and follow it to a neighboring site. Find the line that this new line is paired to on that site, and follow that line to a neighboring site, etc. Do this until you complete a loop. Now, find the highest remaining spin, call it  $\sigma_q$, which will also come out of a full site. Repeat the above process to get another loop. Make loops until you run out of lines. Let $l$ be the number of loops, and $x$ be the number of full sites in the diagram. Insert a factor of $m^l(-1)^{x-l}$.
\end{itemize}
\subsection{Derivation of sign and spin sum rule\label{spinsign}}

Rules 1-4 follow from the expression for $\frac{Z}{Z_0}$ by inspection. We now derive rule 5. We will first show why it works for some examples and then prove that it works for all diagrams. The examples will also illustrate the other rules.
\subsubsection{Examples}
Consider the following example from fourth order.
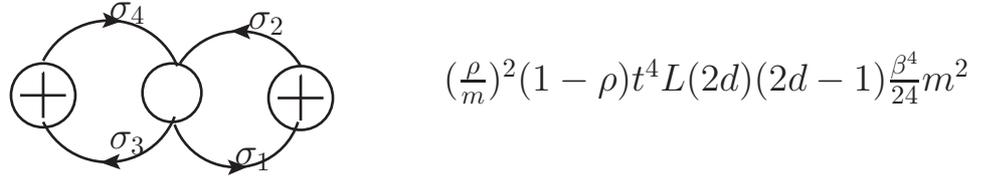
\begin{center}
\fcolorbox{white}{white}{
  \begin{picture}(198,79) (36,1)
    \SetWidth{1.0}
    \SetColor{Black}
    \COval(49,32)(12,12)(90.0){Black}{White}\Line(40.515,32)(57.485,32)\Line(49,40.485)(49,23.515)
    \COval(97,33)(11,11)(0){Black}{White}
    \COval(145,31)(12,12)(90.0){Black}{White}\Line(136.515,31)(153.485,31)\Line(145,39.485)(145,22.515)
    \Arc[arrow,arrowpos=0.5,arrowlength=5,arrowwidth=2,arrowinset=0.2,clock](73.545,32.619)(27.49,153.233,22.186)
    \Arc[arrow,arrowpos=0.5,arrowlength=5,arrowwidth=2,arrowinset=0.2,clock](73.065,33.662)(26.742,-21.18,-154.145)
    \Arc[arrow,arrowpos=0.5,arrowlength=5,arrowwidth=2,arrowinset=0.2](122.193,29.679)(26.413,30.289,147.165)
    \Arc[arrow,arrowpos=0.5,arrowlength=5,arrowwidth=2,arrowinset=0.2](120.831,29.288)(24.288,-160.048,-17.461)
    \Text(74,59)[lb]{\Large{\Black{$\sigma_4$}}}
    \Text(74,10)[lb]{\Large{\Black{$\sigma_3$}}}
    \Text(126,55)[lb]{\Large{\Black{$\sigma_2$}}}
    \Text(121,4)[lb]{\Large{\Black{$\sigma_1$}}}
    \Text(199,29)[lb]{\Large{\Black{$(\frac{\rho}{m})^2(1-\rho)t^4L(2d)(2d-1)\frac{\beta^4}{24}m^2$}}}
  \end{picture}
}
\end{center}
The original ordering of operators in the expectation value is $\langle X_{j_1'}^{\sigma_10}X_{j_1}^{0\sigma_1}X_{j_2'}^{\sigma_20}X_{j_2}^{0\sigma_2}X_{j_3'}^{\sigma_30}X_{j_3}^{0\sigma_3}X_{j_4'}^{\sigma_40}X_{j_4}^{0\sigma_4}\rangle_0$ Let us label the diagram by writing the indices of the operators acting on a given site above the site.
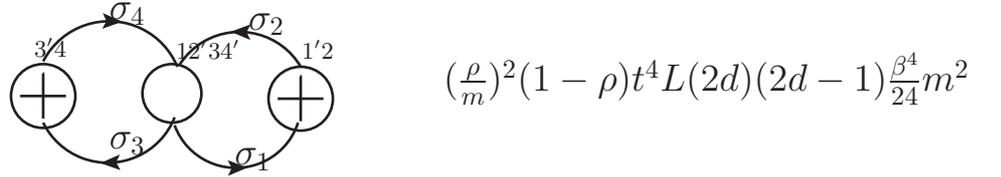
\begin{center}
\fcolorbox{white}{white}{
  \begin{picture}(198,79) (36,1)
    \SetWidth{1.0}
    \SetColor{Black}
    \COval(49,32)(12,12)(90.0){Black}{White}\Line(40.515,32)(57.485,32)\Line(49,40.485)(49,23.515)
    \COval(97,33)(11,11)(0){Black}{White}
    \COval(145,31)(12,12)(90.0){Black}{White}\Line(136.515,31)(153.485,31)\Line(145,39.485)(145,22.515)
    \Arc[arrow,arrowpos=0.5,arrowlength=5,arrowwidth=2,arrowinset=0.2,clock](73.545,32.619)(27.49,153.233,22.186)
    \Arc[arrow,arrowpos=0.5,arrowlength=5,arrowwidth=2,arrowinset=0.2,clock](73.065,33.662)(26.742,-21.18,-154.145)
    \Arc[arrow,arrowpos=0.5,arrowlength=5,arrowwidth=2,arrowinset=0.2](122.193,29.679)(26.413,30.289,147.165)
    \Arc[arrow,arrowpos=0.5,arrowlength=5,arrowwidth=2,arrowinset=0.2](120.831,29.288)(24.288,-160.048,-17.461)
    \Text(74,59)[lb]{\Large{\Black{$\sigma_4$}}}
    \Text(74,10)[lb]{\Large{\Black{$\sigma_3$}}}
    \Text(126,55)[lb]{\Large{\Black{$\sigma_2$}}}
    \Text(121,4)[lb]{\Large{\Black{$\sigma_1$}}}
    \Text(199,29)[lb]{\Large{\Black{$(\frac{\rho}{m})^2(1-\rho)t^4L(2d)(2d-1)\frac{\beta^4}{24}m^2$}}}
    \Text(46,47)[lb]{\small{\Black{$3'4$}}}
    \Text(99,46)[lb]{\small{\Black{$12'34'$}}}
    \Text(146,46)[lb]{\small{\Black{$1'2$}}}
  \end{picture}
}
\end{center}
Let us now determine the sign of the diagram. This is the number of transpositions it takes to get between the following 2 orderings: 3'412'34'1'2 and 1'12'23'34'4. We can move a number past 2 numbers without changing the sign. We can also move a pair of numbers anywhere we want without changing the sign. Let us pair the numbers as indicated in rule 5 above: (3',4)(1,2')(3,4')(1',2). Let us now move the (3,4') pair inside the (3',4) pair to obtain (3'34'4)(1,2')(1',2). Let us now move the (1,2') pair inside the (1',2) pair to obtain (3'34'4)(1'12'2). Now we can move x'x pairs to get the desired order 1'12'23'34'4. Hence, the sign of the diagram is (+). Now we determine the contribution from the spin sum. For empty sites, we must first create a particle. For filled sites, we must first destroy a particle. Hence, recalling Eqs. (\ref{fulldelta}) and (\ref{emptydelta}), we see that the pairings in rule 5 correspond to the Kronecker deltas in these formulas. In addition, x' and x share the same spin label $\sigma_x$. Hence, in our expression (3'34'4)(1'12'2), all numbers within a given parenthesis share the same spin. Hence, the spin sum is $m^2$. 

Alternatively, if we go back to the diagram and make loops by the process indicated in step 5 of the rules, we will see that the first loop corresponds to (3'34'4), while the second one corresponds to (1'12'2). Since there are 2 full sites in the diagram and 2 loops, rule 5 says that the sign and spin sum give a factor of $(-1)^{2-2}m^2=+ \ m^2$, which is exactly what we found. Using the other rules as well, the contribution of the diagram can be found and is written next to the diagram. We have assumed that the lattice is a d-dimensional hypercube. Let us consider another example.
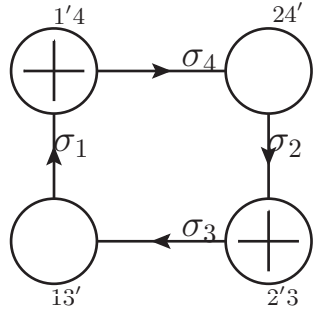
\begin{center}
\fcolorbox{white}{white}{
  \begin{picture}(228,121) (63,-8)
    \SetWidth{1.0}
    \SetColor{Black}
    \COval(80,73)(16,16)(90.0){Black}{White}\Line(68.686,73)(91.314,73)\Line(80,84.314)(80,61.686)
    \COval(80,9)(16,16)(0){Black}{White}
    \COval(160,9)(16,16)(90.0){Black}{White}\Line(148.686,9)(171.314,9)\Line(160,20.314)(160,-2.314)
    \COval(160,73)(16,16)(0){Black}{White}
    \Line[arrow,arrowpos=0.5,arrowlength=5,arrowwidth=2,arrowinset=0.2](96,73)(144,73)
    \Line[arrow,arrowpos=0.5,arrowlength=5,arrowwidth=2,arrowinset=0.2](144,9)(96,9)
    \Line[arrow,arrowpos=0.5,arrowlength=5,arrowwidth=2,arrowinset=0.2](160,57)(160,25)
    \Line[arrow,arrowpos=0.5,arrowlength=5,arrowwidth=2,arrowinset=0.2](80,25)(80,57)
    \Text(128,73)[lb]{\Large{\Black{$\sigma_4$}}}
    \Text(160,41)[lb]{\Large{\Black{$\sigma_2$}}}
    \Text(128,9)[lb]{\Large{\Black{$\sigma_3$}}}
    \Text(80,41)[lb]{\Large{\Black{$\sigma_1$}}}
    \Text(230,57)[lb]{\Large{\Black{$(-1)(\frac{\rho}{m})^2(1-\rho)^2t^4L(2d)(2d-2)\frac{\beta^4}{24}m$}}}
    \Text(80,91)[lb]{\small{\Black{$1'4$}}}
    \Text(162,92)[lb]{\small{\Black{$24'$}}}
    \Text(160,-15)[lb]{\small{\Black{$2'3$}}}
    \Text(79,-15)[lb]{\small{\Black{$13'$}}}
  \end{picture}
}
\end{center}
The initial ordering is 1'424'2'313'. Again, pair the numbers according to rule 5: (1',4)(2,4')(2',3)(1,3'). Move the (1,3') pair into (1',4) pair to obtain: (1'13'4)(2,4')(2',3). Now move the (2',3) pair inside the loop to obtain: (1'13'2'34)(2,4'). Now we need a transposition to get 3 to the right of 3'. This gives a (-) sign. We obtain: -(1'13'32'4)(2,4'). Now put (2,4') into the loop to obtain: -(1'1'3'32'24'4) $=$ -(1'1'2'23'34'4). The reason that we got an overall (-) sign for this diagram was because the (2',3) pair was in the ``wrong" order with the primed number to the left of the unprimed one instead of the other way around. Both (1,3') and (2,4') were in the ``right" order and hence generated no (-) sign. The wrong order came about because (2',3) was on a full site rather than an empty site. The full site pair (1',4) did not generate a (-) sign because it started the loop. Hence, full sites generate (-) signs except when they start loops. This is the reason for the factor $(-1)^{x-l}$ in rule 5. Since we put everything into one parenthesis this time, the spin sum gives a factor of m. 

Alternatively, using step 5 of the rules we would break the diagram into one loop. In addition, it has 2 full sites. So the factor from the sign and the spin sum should be $(-1)^{2-1}m^1=- \ m$ as we have already found. Using the other rules as well, the contribution of the diagram can be found and is written next to the diagram.

The only case we have yet to consider is when full sites have multiple pairs on them. Let us consider a full site which has the numbers $x'_1x_2x'_3x_4\ldots x'_{p-1}x_p$ on it. In this case, the pairings should be made according to rule 5: $(x'_1,x_p)(x_2,x'_3)\ldots(x_{p-2},x'_{p-1})$. Making the pairings in this way does not generate a (-) sign because $x_p$ has to move past a certain number of pairs to get to the right of $x'_1$. We see that only the  $(x'_1,x_p)$ pair is in the ``wrong" order while the other pairs are in the ``right" order. Hence, only this pair has the potential to generate a (-) sign and will do so unless it is used to start a loop. Let us consider a concrete example from sixth order which illustrates this.
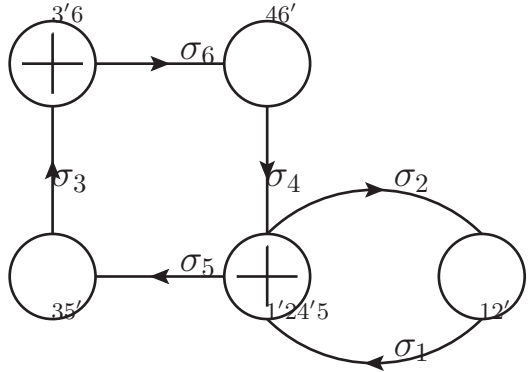
\begin{center}
\fcolorbox{white}{white}{
  \begin{picture}(292,156) (95,-43)
    \SetWidth{1.0}
    \SetColor{Black}
    \COval(112,76)(16,16)(90.0){Black}{White}\Line(100.686,76)(123.314,76)\Line(112,87.314)(112,64.686)
    \COval(192,-4)(16,16)(90.0){Black}{White}\Line(180.686,-4)(203.314,-4)\Line(192,7.314)(192,-15.314)
    \COval(112,-4)(16,16)(0){Black}{White}
    \COval(192,76)(16,16)(0){Black}{White}
    \COval(272,-4)(16,16)(0){Black}{White}
    \Line[arrow,arrowpos=0.5,arrowlength=5,arrowwidth=2,arrowinset=0.2](128,76)(176,76)
    \Line[arrow,arrowpos=0.5,arrowlength=5,arrowwidth=2,arrowinset=0.2](192,60)(192,12)
    \Line[arrow,arrowpos=0.5,arrowlength=5,arrowwidth=2,arrowinset=0.2](176,-4)(128,-4)
    \Line[arrow,arrowpos=0.5,arrowlength=5,arrowwidth=2,arrowinset=0.2](112,12)(112,60)
    \Arc[arrow,arrowpos=0.5,arrowlength=5,arrowwidth=2,arrowinset=0.2,clock](232,-28)(56.569,135,45)
    \Arc[arrow,arrowpos=0.5,arrowlength=5,arrowwidth=2,arrowinset=0.2,clock](232,20)(56.569,-45,-135)
    \Text(160,76)[lb]{\Large{\Black{$\sigma_6$}}}
    \Text(192,28)[lb]{\Large{\Black{$\sigma_4$}}}
    \Text(160,-4)[lb]{\Large{\Black{$\sigma_5$}}}
    \Text(112,28)[lb]{\Large{\Black{$\sigma_3$}}}
    \Text(240,28)[lb]{\Large{\Black{$\sigma_2$}}}
    \Text(240,-36)[lb]{\Large{\Black{$\sigma_1$}}}
    \Text(112,92)[lb]{\small{\Black{$3'6$}}}
    \Text(192,92)[lb]{\small{\Black{$46'$}}}
    \Text(192,-20)[lb]{\small{\Black{$1'24'5$}}}
    \Text(112,-20)[lb]{\small{\Black{$35'$}}}
    \Text(272,-20)[lb]{\small{\Black{$12'$}}}
    \Text(300,28)[lb]{\Large{\Black{$(\frac{\rho}{m})^2(1-\rho)^3t^6L(2d)(2d-2)^2\frac{\beta^6}{6!}(-1)m$}}}
  \end{picture}
}
\end{center}
The initial ordering is 3'646'12'1'24'535'. Now we pair the numbers: (3',6)(4,6')(1,2')(1',5)(2,4')(3,5'), and perform the necessary steps to bring them into the desired order.
\begin{eqnarray} && (3',6)(4,6')(1,2')(1',5)(2,4')(3,5')=(3'35'6)(4,6')(1,2')(1',5)(2,4')\nonumber\\
&& =(3'35'1'56)(4,6')(1,2')(2,4')=-(3'35'51'6)(4,6')(1,2')(2,4')\nonumber\\
&&=-(3'35'51'12'6)(4,6')(2,4')=-(3'35'51'12'24'6)(4,6')\nonumber\\
&&=-(3'35'51'12'24'46'6)=-(1'12'23'34'45'56'6).
\end{eqnarray}
Note that the full site on the bottom right corner of the square generated a (-) sign from its ``wrong" pair (1',5), which was not used to start a loop. The ``wrong" pair (3',6) on the full site at the top left corner of the square was used to start a loop, and hence did not generate a (-) sign. Therefore, the overall sign of the diagram is (-). All of the pairs were put into one parenthesis, and hence the spin sum gave a factor of $m$. 

Alternatively, we could use rule 5 according to which we find that the diagram has 1 loop. We also see that it has 2 full sites. Hence, the sign and spin sum give the contribution $(-1)^{2-1}(m)^1=-m$, which matches our previous result. The total contribution of the diagram is again written next to the diagram.

\subsubsection{Proof of the general case \label{proofspinsign}}
We can now write down a rigorous proof for rule 5 for an $n^{th}$ order diagram. For a given vertex, we indicate an incoming line with a prime, and an outgoing line with no prime. In the notation of rule 5, empty sites have the pairs $(x_1,x_2')(x_3,x_4')\ldots(x_{p-1},x_{p}')$ on them, while full sites have the pairs $(x_1',x_p)(x_2,x_3')\ldots(x_{p-2},x_{p-1}')$ on them. The loop starts with the pair $(x',n)$ on a full site. This corresponds to the $(x_1',x_p)$ pair on this site. The next pair in this loop is either $(x,y')$ or $(y',x)$, where the latter can only correspond to the $(x_1',x_p)$ pair on a full site. The loop now becomes $\pm (x'xy'n)$ where $(x,y')$ yields $+$ and $(y',x)$ yields $-$. This process continues until we come across $n'$ at which point we complete the first loop.

Suppose that the first loop has not exhausted all of the lines in the diagram. Of the lines not used in the first loop, find the one with the highest spin label, $\sigma_q$. We now show that $q$ must occur in a pair of the form $(u',q)$, corresponding to the $(x_1',x_p)$ pair on a full site. The number $q$ must be on some vertex. Remove the pairs that were used in the first loop from this vertex. $q$ must be the biggest number of those still left on this vertex. However, we see that this could only occur if $x_p=q$ for a full site in the original diagram. 

The pair $(u',q)$ will start the second loop, which will be formed in exactly the same way as the first loop. We repeat the same reasoning to show that all of the loops are started by pairs of the form $(x_1',x_p)$ on full sites. Those $(x_1',x_p)$ pairs on full sites which do not start loops will generate (-) signs. This means that of the $x$ full sites, $x-l$ generate $(-)$ signs. Hence, the overall sign of the diagram is $(-1)^{x-l}$. Furthermore, the lines in a loop must share the same spin, independent of the other loops. Hence, the spin sum is $m^l$. Thus, the overall factor from the sign and spin sum is $(-1)^{x-l}m^l$ as stated in rule 5.                                        

\subsection{Loss and recovery of the linked cluster theorem \lab{linked-cluster}}

In the case of Feynman diagrams or the Meyer cluster expansion of the classical gas, disconnected diagrams arise in the expression for the partition function, but are eliminated from the thermodynamic potential upon taking the log of the partition function. This so-called ``linked cluster theorem" \cite{Nozieres,Domb} is a consequence of two properties of the disconnected diagrams. a) The contribution of a disconnected diagram is the product of the contributions of the connected components from which it is formed. b) The combinatorial factors involved in permuting the labels work out in just the right way for the cancellation to occur. In the case of the expansion at hand, property b) continues to hold. However, property a) breaks down due to the restricted lattice sum in which the distinct vertices of the diagram represent distinct sites on the lattice. This is illustrated below by the simplest disconnected diagrams in the calculation of $\frac{Z}{Z_0}$, which appear in 4th order. 
\begin{center}
\fcolorbox{white}{white}{
  \begin{picture}(387,99) (100,0)
    \SetWidth{0.5}
    \SetColor{Black}
    \Text(128,67)[lb]{\Large{\Black{$\sigma_4$}}}
    \Text(128,3)[lb]{\Large{\Black{$\sigma_3$}}}
    \SetWidth{1.0}
    \COval(96,35)(16,16)(90.0){Black}{White}\Line(84.686,35)(107.314,35)\Line(96,46.314)(96,23.686)
    \COval(160,35)(16,16)(0){Black}{White}

    \Arc[arrow,arrowpos=0.5,arrowlength=5,arrowwidth=2,arrowinset=0.2,clock](128,27)(40,143.13,36.87)
    \Arc[arrow,arrowpos=0.5,arrowlength=5,arrowwidth=2,arrowinset=0.2,clock](128,43)(40,-36.87,-143.13)
    \COval(96,35)(16,16)(90.0){Black}{White}\Line(84.686,35)(107.314,35)\Line(96,46.314)(96,23.686)
    \COval(160,35)(16,16)(0){Black}{White}

    \Arc[arrow,arrowpos=0.5,arrowlength=5,arrowwidth=2,arrowinset=0.2,clock](128,27)(40,143.13,36.87)
    \Arc[arrow,arrowpos=0.5,arrowlength=5,arrowwidth=2,arrowinset=0.2,clock](128,43)(40,-36.87,-143.13)
    \COval(221,37)(16,16)(90.0){Black}{White}\Line(209.686,37)(232.314,37)\Line(221,48.314)(221,25.686)
    \COval(285,37)(16,16)(0){Black}{White}

    \Arc[arrow,arrowpos=0.5,arrowlength=5,arrowwidth=2,arrowinset=0.2,clock](253,29)(40,143.13,36.87)
    \Arc[arrow,arrowpos=0.5,arrowlength=5,arrowwidth=2,arrowinset=0.2,clock](253,45)(40,-36.87,-143.13)
    \Text(254,70)[lb]{\Large{\Black{$\sigma_2$}}}
    \Text(254,7)[lb]{\Large{\Black{$\sigma_1$}}}
    \Text(310,37)[lb]{\Large{\Black{$\rho^2(1-\rho)^2\frac{\beta^4}{24}t^4L(2d)[2dL-4(2d-1)-2]$}}}
  \end{picture}
}
\end{center}
\begin{center}
\fcolorbox{white}{white}{
  \begin{picture}(387,99) (100,0)
    \SetWidth{0.5}
    \SetColor{Black}
    \Text(128,67)[lb]{\Large{\Black{$\sigma_4$}}}
    \Text(128,3)[lb]{\Large{\Black{$\sigma_2$}}}
    \SetWidth{1.0}
    \COval(96,35)(16,16)(90.0){Black}{White}\Line(84.686,35)(107.314,35)\Line(96,46.314)(96,23.686)
    \COval(160,35)(16,16)(0){Black}{White}

    \Arc[arrow,arrowpos=0.5,arrowlength=5,arrowwidth=2,arrowinset=0.2,clock](128,27)(40,143.13,36.87)
    \Arc[arrow,arrowpos=0.5,arrowlength=5,arrowwidth=2,arrowinset=0.2,clock](128,43)(40,-36.87,-143.13)
    \COval(96,35)(16,16)(90.0){Black}{White}\Line(84.686,35)(107.314,35)\Line(96,46.314)(96,23.686)
    \COval(160,35)(16,16)(0){Black}{White}

    \Arc[arrow,arrowpos=0.5,arrowlength=5,arrowwidth=2,arrowinset=0.2,clock](128,27)(40,143.13,36.87)
    \Arc[arrow,arrowpos=0.5,arrowlength=5,arrowwidth=2,arrowinset=0.2,clock](128,43)(40,-36.87,-143.13)
    \COval(221,37)(16,16)(90.0){Black}{White}\Line(209.686,37)(232.314,37)\Line(221,48.314)(221,25.686)
    \COval(285,37)(16,16)(0){Black}{White}

    \Arc[arrow,arrowpos=0.5,arrowlength=5,arrowwidth=2,arrowinset=0.2,clock](253,29)(40,143.13,36.87)
    \Arc[arrow,arrowpos=0.5,arrowlength=5,arrowwidth=2,arrowinset=0.2,clock](253,45)(40,-36.87,-143.13)
    \Text(254,70)[lb]{\Large{\Black{$\sigma_3$}}}
    \Text(254,7)[lb]{\Large{\Black{$\sigma_1$}}}
    \Text(310,37)[lb]{\Large{\Black{$\rho^2(1-\rho)^2\frac{\beta^4}{24}t^4L(2d)[2dL-4(2d-1)-2]$}}}
  \end{picture}
}
\end{center}
\begin{center}
\fcolorbox{white}{white}{
  \begin{picture}(387,99) (100,0)
    \SetWidth{0.5}
    \SetColor{Black}
    \Text(128,67)[lb]{\Large{\Black{$\sigma_4$}}}
    \Text(128,3)[lb]{\Large{\Black{$\sigma_1$}}}
    \SetWidth{1.0}
    \COval(96,35)(16,16)(90.0){Black}{White}\Line(84.686,35)(107.314,35)\Line(96,46.314)(96,23.686)
    \COval(160,35)(16,16)(0){Black}{White}

    \Arc[arrow,arrowpos=0.5,arrowlength=5,arrowwidth=2,arrowinset=0.2,clock](128,27)(40,143.13,36.87)
    \Arc[arrow,arrowpos=0.5,arrowlength=5,arrowwidth=2,arrowinset=0.2,clock](128,43)(40,-36.87,-143.13)
    \COval(96,35)(16,16)(90.0){Black}{White}\Line(84.686,35)(107.314,35)\Line(96,46.314)(96,23.686)
    \COval(160,35)(16,16)(0){Black}{White}

    \Arc[arrow,arrowpos=0.5,arrowlength=5,arrowwidth=2,arrowinset=0.2,clock](128,27)(40,143.13,36.87)
    \Arc[arrow,arrowpos=0.5,arrowlength=5,arrowwidth=2,arrowinset=0.2,clock](128,43)(40,-36.87,-143.13)
    \COval(221,37)(16,16)(90.0){Black}{White}\Line(209.686,37)(232.314,37)\Line(221,48.314)(221,25.686)
    \COval(285,37)(16,16)(0){Black}{White}

    \Arc[arrow,arrowpos=0.5,arrowlength=5,arrowwidth=2,arrowinset=0.2,clock](253,29)(40,143.13,36.87)
    \Arc[arrow,arrowpos=0.5,arrowlength=5,arrowwidth=2,arrowinset=0.2,clock](253,45)(40,-36.87,-143.13)
    \Text(254,70)[lb]{\Large{\Black{$\sigma_3$}}}
    \Text(254,7)[lb]{\Large{\Black{$\sigma_2$}}}
    \Text(310,37)[lb]{\Large{\Black{$\rho^2(1-\rho)^2\frac{\beta^4}{24}t^4L(2d)[2dL-4(2d-1)-2]$}}}
  \end{picture}
}
\end{center}

Applying the rules for $\frac{Z}{Z_0}$, we see that they all have an identical contribution which is written next to each of the diagrams. The permutation of line labels leads to $\binom{4}{2}\frac{1}{2!}=3$ diagrams. There are 4 lines, 2 of which must be chosen for the connected component on the left. However, since the 2 components are identical, exchanging all of the labels between them does not lead to a new labeling. 

The lattice sum for each of these disconnected diagrams is $(2dL)[2dL-4(2d-1)-2]$. This lattice sum comes about because the vertices are restricted to being distinct sites on the lattice. For a second, let us suppose that this is not the case, and that the lattice sum is unrestricted. The lattice sum would then simply be $(2dL)^2$. In this case, the total contribution from the three disconnected diagrams would be $(\frac{Z}{Z_0})^{(4)}_{disconnected} =\rho^2(1-\rho)^2\frac{\beta^4}{8}t^4(2dL)^2$. We relate this contribution to the contribution of the only second order diagram, from which these disconnected diagram are formed.
\begin{center}
\fcolorbox{white}{white}{
  \begin{picture}(275,99) (0,0)
    \SetWidth{1.0}
    \SetColor{Black}
    \COval(96,35)(16,16)(90.0){Black}{White}\Line(84.686,35)(107.314,35)\Line(96,46.314)(96,23.686)
    \COval(160,35)(16,16)(0){Black}{White}

    \Arc[arrow,arrowpos=0.5,arrowlength=5,arrowwidth=2,arrowinset=0.2,clock](128,27)(40,143.13,36.87)
    \Arc[arrow,arrowpos=0.5,arrowlength=5,arrowwidth=2,arrowinset=0.2,clock](128,43)(40,-36.87,-143.13)
    \Text(128,67)[lb]{\Large{\Black{$\sigma_2$}}}
    \Text(128,3)[lb]{\Large{\Black{$\sigma_1$}}}
    \Text(200,16)[lb]{\Large{\Black{$\rho(1-\rho)\frac{\beta^2}{2}t^2L(2d)$}}}
  \end{picture}
}
\end{center}
We find that $(\frac{Z}{Z_0})^{(4)}_{disconnected} = \frac{[(\frac{Z}{Z_0})^{(2)}]^2}{2!}$. This is exactly the factor we need for the linked-cluster theorem to work. 

Let us return to the actual situation, in which the vertices are in fact restricted to being distinct sites on the lattice. In this case, the lattice sum of each diagram is $(2dL)[2dL-4(2d-1)-2]$. This lattice sum contains a term proportional to $L$ and one proportional to $L^2$. However, we have just shown that the one proportional to $L^2$ is cancelled upon taking the log of the partition function, leaving only the one proportional to $L$. We expect this to be the case since the thermodynamic potential is an extensive quantity.

Therefore, we see that the linked cluster theorem may yet be possible, but with a generalized definition of the ``connected components" which go into making a diagram. These generalized connected components will involve overlappings of the original connected components. A disconnected diagram for the partition function involving a number of these generalized connected components will in fact now satisfy both properties a) and b), necessary for the linked cluster theorem to work. Taking the log of the partition function will eliminate all disconnected diagrams leaving only the generalized connected components. This will also provide a rigorous proof for the observation just made, that the thermodynamic potential corresponds to the term proportional to $L$ in the partition function. The proof of the linked cluster theorem in terms of generalized connected components will hinge upon a formula for the restricted lattice sum of a disconnected diagram with $n$ of the original components, which we shall now derive.
\subsection{Formula for the restricted lattice sum of disconnected diagrams \label{lsdisc}}
\subsubsection{Restricted lattice sum of disconnected diagrams with 2 or 3 components}
The simplest instance of a disconnected diagram is one of the 4th order disconnected diagrams considered above. One of the connected components can be placed anywhere on the lattice, hence the factor $(2dL)$. The other component can be placed anywhere on the lattice such that none of its sites overlap any of the sites of the first one. The number of ways it can overlap the first one with just one site is $4(2d-1)$. The number of ways it can overlap it with both sites is $2$. Hence the factor $[2dL-4(2d-1)-2]$. The lattice sum is thus $(2dL)[2dL-4(2d-1)-2]$. 

In the case where there are more than two disconnected components, it will be difficult to calculate the lattice sum by adding on one component at a time, because for the third one, its options depend on how far apart the first two are on the lattice. Hence, we need a systematic way of calculating the lattice sums. 
Consider a disconnected diagram with 2 components.
\beq LS[A  \ dc \ B] = LS[AB] - LS[(A\cap B)]. \eeq
Here, $LS[A \  dc \ B]$ indicates the lattice sum of A disconnected from B, and is what we are trying to calculate for a disconnected diagram. $LS[AB]$ indicates the lattice sum where $A$ and $B$ ignore each other, and can each be placed anywhere on the lattice. Hence, $LS[AB]=LS[A]LS[B]$. $LS[(A\cap B)]$ indicates the lattice sum where $A$ and $B$ somehow overlap. We generalize this notation. $LS[D_1 \ dc \ D_2 \ dc  \ldots dc  \ D_n]$ indicates the lattice sum of the components $D_1 \ldots D_n$ in which they are not allowed to overlap each other in any way, and is the object that we need a systematic way of calculating. $LS[D_1D_2\ldots D_n]=LS[D_1]LS[D_2]\ldots LS[D_n]$. Finally, $LS[(D_1\cap D_2\cap \ldots \cap D_n)]$ indicates the lattice sum of the components $D_1 \ldots D_n$ in which they must overlap to form one connected component, but no two of the $D_1 \ldots D_n$ are required to overlap each other. 

Consider a disconnected diagram with 3 components.
\beq LS[A \ dc \ B \ dc \ C] = LS[ABC]-LS[(A\cap B\cap C)] - LS[(A\cap B) \ dc \ C] - LS[(A\cap C) \ dc \ B]-LS[(B\cap C) \ dc \ A]. \label{AdBdC} \eeq     
Here, $LS[(A\cap B) \ dc \ C]$ indicates the lattice sum where $A$ and $B$ overlap to form a connected component, which is then not allowed to overlap the component $C$. This formula states that to obtain the lattice sum of $A$, $B$, and $C$ not overlapping in any way, we take the lattice sum of them ignoring each other, and subtract the lattice sums of all of the possible ways in which they can overlap (either overlapping to form one connected component or two, which are then not allowed to overlap each other). According to the formula for two components, we have
\beq LS[(A\cap B)  \ dc \  C] = LS[(A\cap B) C] - LS[((A\cap B)\cap C)].\label{AiBdC} \eeq
Here, $LS[((A\cap B)\cap C)]$ indicates the lattice sum in which first $A$ and $B$ must overlap to form a connected component, and then the resulting connected component must overlap $C$. This is not the same as the term $LS[(A\cap B\cap C)]$ in which $A$, $B$, and $C$ must overlap to form a connected component, but may do so without $A$ overlapping $B$. In general, we use the notation $(D_1\cap D_2 \cap \ldots \cap D_n)$ to indicate that first $D_1\ldots D_n$ must overlap to form a connected component. Combining \disp{AdBdC} with \disp{AiBdC}, we see that
\begin{eqnarray} &&LS[A \ dc \ B \ dc \ C] = LS[ABC]-LS[(A\cap B\cap C)] - LS[(A\cap B) C]+ LS[((A\cap B)\cap C)]\nonumber\\
&& - LS[(A\cap C) B]+ LS[((A\cap C)\cap B)]- LS[(B\cap C) A]+ LS[((B\cap C)\cap A)].\label{3disc} 
\end{eqnarray}
\subsubsection{Restricted lattice sum of disconnected diagrams with n components}
Consider a disconnected diagram with n components $D_1\ldots D_n$. The object we wish to calculate is $LS[D_1 \ dc \ D_2  \ dc\ldots dc \ D_n]$. We think of each term in \disp{3disc} as coming from a particular ``configuration". For example, the term $LS[((B\cap C)\cap A)]$ comes from the configuration $((B\cap C)\cap A)$. We call $D_1D_2 \ldots D_n$ the initial configuration.
$LS[D_1 \ dc \ D_2 \ dc \ldots dc  \ D_n]$ is a sum of terms, which are generated by the following set of rules:
\begin{itemize}
\item[1] Starting from the initial configuration, arbitrarily group the components $D_1 \ldots D_n$. Enclose each group with a parenthesis and place intersection symbols between the members of a single group. For a component not grouped with any of the other ones we imagine that there is a parenthesis around it but we do not draw it in.
\item[2] Identify the {\bf outer parentheses}. These are the parentheses not enclosed in any other parenthesis. Arbitrarily group the outer parentheses. Enclose each group with a parenthesis and place intersection symbols between the members of a single group.
\item[3] We denote each time you group objects as a step. Take anywhere from $0$ to $n-1$ steps, each time grouping the outer parentheses, to get from the initial configuration to the final configuration associated with this sequence of steps. The final configuration will have $p\leq n$ outer parentheses. Each outer parenthesis represents an overlapping of some subset of the components present in the initial configuration. {\bf Let us number the outer parentheses by the index i, and denote the overlapping of components represented by each outer parenthesis as $w_i$.} Let $s$ denote the number of steps taken. Then, the contribution of this sequence of steps to $LS[D_1 \ dc \ D_2 \ dc\ldots dc \  D_n]$ is given by $(-1)^sLS[w_1w_2\ldots w_p ]=(-1)^sLS[w_1]LS[w_2]\ldots LS[w_p ]$.
\item[4] $LS[D_1 \ dc \ D_2 \ dc\ldots dc \ D_n]$ is given by the sum of all terms generated by a distinct sequence of steps.
\end{itemize}

We illustrate these rules with a couple of examples from the case $n=3$ (\disp{3disc}).  Consider the configuration $(A\cap B)C$. It is reached from the initial configuration by grouping $A$ and $B$ in the first step. Hence, $s=1$, $w_1 = (A\cap B)$, and $w_2=C$. The contribution of this sequence of steps is therefore $(-1)^1LS[(A\cap B)C]=-LS[A\cap B]LS[C]$. Consider the configuration $((A\cap B)\cap C)$. It is reached from the initial configuration by grouping $A$ and $B$ in the first step. The second step consists of grouping $(A \cap B)$ with $C$. Hence, $s=2$, and $w_1 = ((A\cap B)\cap C)$. The contribution of this sequence of steps is therefore $(-1)^2LS[((A\cap B)\cap C)]=LS[((A\cap B)\cap C)]$.

For $n\leq3$ components, each final configuration must be reached by a unique sequence of steps starting from the initial configuration. However, this is not the case for $n\geq4$. For $n=4$, consider the configuration $(D_1 \cap D_2)(D_3 \cap D_4)$.
There are three distinct sequences of steps to get from the initial configuration to this configuration. One sequence involves only one step in which $D_1$ and $D_2$ are grouped together and $D_3$ and $D_4$ are grouped together. Another sequence involves two steps. In the first step, $D_1$ and $D_2$ are grouped together, while in the second one $D_3$ and $D_4$ are grouped together. The third sequence also involves two steps. In the first step, $D_3$ and $D_4$ are grouped together, while in the second one $D_1$ and $D_2$ are grouped together. Since all of these sequences end in the same final configuration, their contribution differs only in the number of steps it takes to get from the initial configuration to this final configuration. The first sequence involves only one step and hence has a contribution of $-LS[(D_1\cap D_2)]LS[(D_3\cap D_4)]$. The other two sequences each involve 2 steps and hence each have a contribution of $LS[(D_1\cap D_2)]LS[(D_3\cap D_4)]$. Hence, the overall contribution of this configuration is $-LS[(D_1\cap D_2)]LS[(D_3\cap D_4)] +2LS[(D_1\cap D_2)]LS[(D_3\cap D_4)]=LS[(D_1\cap D_2)]LS[(D_3\cap D_4)]$.

\subsubsection{Classification of configurations}
We now want to isolate a particular final configuration, and add all of the contributions from the distinct sequences of steps which lead to this final configuration from the initial configuration. This will give us the contribution of this final configuration. If we can do this for any final configuration, then instead of adding contributions from sequences of steps to determine $LS[D_1 \ dc \ D_2 \ dc\ldots dc \  D_n]$, we can add contributions from final configurations. Let us denote an arbitrary final configuration by $\kappa$. Suppose $\kappa$ has $p$ outer parentheses. Then each distinct sequence of steps leading to $\kappa$ will have the contribution $\pm LS[w_1]LS[w_2]\ldots LS[w_p]$, where the choice of plus or minus depends on how many steps there are in that sequence. Hence, the overall contribution from $\kappa$ will be $C_{\kappa}LS[w_1]LS[w_2]\ldots LS[w_p]$ where $C_{\kappa} = \sum\limits_{sequences}(-1)^{s(sequence)}$, where the sum is over all sequences leading to $\kappa$ from the initial configuration. We want to find $C_{\kappa}$ for all $\kappa$. To this end, we classify the different $\kappa$ into types that will share the same value of $C_{\kappa}$.

First, we classify the parentheses appearing in the various configurations. A type 0 parenthesis is the invisible parenthesis enclosing any one of the $n$ components present in the initial configuration. A type 1 parenthesis is a parenthesis which encloses only type 0 parentheses. A type 2 parenthesis encloses at least 1 type 1 parenthesis and zero or more parentheses of lower type. A type 3 parenthesis encloses at least 1 type 2 parenthesis and zero or more parentheses of lower type. A type $k$ parenthesis encloses at least 1 type $k-1$ parenthesis and zero or more parentheses of lower type. We give some examples to illustrate the different types of parentheses. In the following examples, the very outer parenthesis is of the specified type. Type 1 parenthesis: $(D_1\cap D_2)$. Type 2 parenthesis: $((D_1\cap D_2)\cap D_3)$. Type 3 parenthesis: $ (((D_1\cap D_2)\cap D_3)\cap(D_4\cap D_5))$.

We are now in a position to classify all of the final configurations. A type $m_{(i_1,i_2,\ldots,i_m)}$ configuration is a configuration which has $i_1$ type 1 parentheses,  $i_2$ type 2 parantheses, \ldots $i_m$ type m parentheses. Every possible final configuration falls into one of these types. We shall see that all final configurations of a given type have equal $C_{\kappa}$. For a configuration $\kappa$ of type $m_{(i_1,i_2,\ldots,i_m)}$, we shall denote $C_{\kappa}$ by $C_{m_{(i_1,i_2,\ldots,i_m)}}$. The following are examples of different types of configurations. $1_{(1)}$: $(D_1\cap D_2)D_3D_4$, $1_{(2)}$: $(D_1\cap D_2)(D_3\cap D_4)$, $ 2_{(1,1)}$: $((D_1\cap D_2)\cap D_3\cap D_4)$, $ 2_{(2,1)}$: $((D_1\cap D_2)\cap (D_3\cap D_4))$, $ 2_{(1,2)}$: not possible, $ 2_{(2,2)}$: $((D_1\cap D_2)\cap D_3) ((D_4\cap D_5)\cap D_6)$, $ 3_{(1,1,1)}$: $(((D_1\cap D_2)\cap D_3)\cap D_4)$, etc.
\subsubsection{Calculation of $C_{1_{(i_1)}}$}
We now calculate $C_{m_{(i_1,i_2,\ldots,i_m)}}$ for a few simple cases before stating and proving the general result. We start with $C_0$. This is just the initial configuration, reached by making 0 steps.
\beq C_0=1. \eeq
Consider one of the type $1_{(1)}$ configurations. The only sequence of steps by which one can get to this configuration is to form the single type 1 parenthesis on the first step. 
\beq C_{1_{(1)}}=-1. \eeq
Consider one of the type $1_{(2)}$ configurations. As already discussed, there are 3 distinct sequences of steps by which this configuration can be reached. One consists of forming both type 1 parentheses in one step from the initial configuration, while the other two consist of forming one of the type 1 parentheses as the first step and forming the other one as the second step. 
\beq C_{1_{(2)}}=-1+2\times1 = 1. \eeq

The way in which we shall calculate $C_{1_{(3)}}$ illustrates the way in which we shall calculate $C_{m_{(i_1,i_2,\ldots,i_m)}}$ for all $m$. Consider a particular type $1_{(3)}$ configuration. In all of the sequences of steps which lead to this configuration,  the configuration reached right before the last step in the sequence must be either of type $0$, $1_{(1)}$, or $1_{(2)}$. Hence, we have the following formula for $C_{1_{(3)}}$.
\begin{eqnarray} && C_{1_{(3)}} = C_0(-1) + (\text{number of type }1_{(1)} \text{configurations one step away from type }1_{(3)} \text{configuration})C_{1_{(1)}}(-1) \nonumber\\
&&+ (\text{number of type }1_{(2)} \text{configurations one step away from type }1_{(3)} \text{configuration})C_{1_{(2)}}(-1). \end{eqnarray}
Thus, our strategy in calculating $C_{m_{(i_1,i_2,\ldots,i_m)}}$ is to calculate these coefficients in the correct order, so that by the time we are calculating the coefficient for a particular type of configuration, we have already calculated the coefficients for all configurations that are within one step of it. This makes the initially daunting task of calculating $C_{m_{(i_1,i_2,\ldots,i_m)}}$ very manageable.

Returning to our calculation of $C_{1_{(3)}}$, a type $1_{(3)}$ configuration can be reached in one step from the initial configuration by forming all three of the type 1 parentheses in this single step. It can be reached in one step from $\binom{3}{1}=3$  distinct type $1_{(1)}$ configurations, one for each of the type 1 parentheses that define the  $1_{(3)}$ configuration. This is done by forming the other two type 1 parentheses in that step. It can be reached in one step from $\binom{3}{2}=3$ distinct type $1_{(2)}$ configurations, one for each choice of 2 of the type 1 parentheses that define it. This is done by forming the other type 1 parenthesis in that step. Therefore,
\beq C_{1_{(3)}} = C_0(-1) + 3C_{1_{(1)}}(-1)+3C_{1_{(2)}}(-1)=1(-1)+3(-1)(-1)+3(1)(-1)=-1.  \eeq 
Similarly, we can calculate $C_{1_{(4)}}$.
\beq C_{1_{(4)}} = -1+\binom{4}{1}(-1)(-1)+\binom{4}{2}(1)(-1)+\binom{4}{3}(-1)(-1)=-1+4-6+4=8-7=1. \eeq
We can now calculate $C_{1_{(i_1)}}$.
\beq C_{1_{(i_1)}} = -1 +\binom{i_1}{1}-\binom{i_1}{2}+\binom{i_1}{3}-\ldots+(-1)^{i_1}\binom{i_1}{i_1-1}=(-1)^{i_1}. \eeq
\subsubsection{Calculation of $C_{m_{(i_1,i_2,\ldots,i_m)}}$ for all m}
We are now ready to state and prove the general formula for $C_{m_{(i_1,i_2,\ldots,i_m)}}$
\beq C_{m_{(i_1,i_2,\ldots,i_m)}} = (-1)^{i_1+i_2+\ldots +i_m}. \eeq

We prove this formula by induction. Before stating the inductive hypothesis, we order the coefficients $C_{m_{(i_1,i_2,\ldots,i_m)}}$. To compare the coefficients $C_{m_{(i_1,i_2,\ldots,i_m)}}$ and $C_{k_{(j_1,j_2,\ldots,j_k)}}$, find the leftmost entry where they differ, where from left to right the entries are $m, i_1,\ldots ,i_m$ and $k,j_1,\ldots,j_k$. The one that has the bigger number in this entry is greater according to this ordering. We write the coefficients in order from least to greatest: 
\beq C_0,C_{1_{(1)}},C_{1_{(2)}},\ldots,C_{2_{(1,1)}},C_{2_{(2,1)}},C_{2_{(2,2)}},C_{2_{(3,1)}},C_{2_{(3,2)}},C_{2_{(3,3)}},\ldots ,C_{3_{(1,1,1)}},\ldots ,C_{n-1_{(1,1,1,\ldots,1)}}. \eeq
{\bf Inductive hypothesis}: $C_{m_{(i_1,i_2,\ldots,i_m)}} = (-1)^{i_1+i_2+\ldots +i_m}$ holds for all coefficients less than or equal to the $r^{th}$ coefficient in the above sequence of coefficients. We have already proven the base case, so we now prove the inductive step. 

Suppose that the $r + 1^{st}$ coefficient is $C_{m_{(i_1,i_2,\ldots,i_m)}}$. The inductive hypothesis implies that $C_{k_{(j_1,j_2,\ldots,j_k)}}=(-1)^{j_1+j_2+\ldots +j_k}$ if all of the inequalities $k\leq m$, $j_1\leq i_1$, $j_2\leq i_2$, \ldots , $j_k\leq i_k$ are satisfied, except for the case where all of the inequalities are equalities. We now split the $i_1$ type 1 parentheses into 2 sets, $\alpha$ and $\beta$. In the $\alpha$ set, we put those type 1 parentheses which are enclosed by other (higher type) parentheses. We put the rest into the set $\beta$. We do the same for the type 2 parentheses, type 3 parentheses, $\ldots$ type $m-1$ parentheses. All of the type m parentheses are put into the set $\beta$ since they can't be enclosed by other parentheses. Consider the configurations from which the type  $m_{(i_1,i_2,\ldots,i_m)}$ configuration in question can be reached in one step. Any such configuration must have all of the parentheses in the set $\alpha$, since these are enclosed by higher type parentheses, but we can only make one more step. In addition to these, it can have anywhere from zero to all but one of  the parentheses from the set $\beta$, since any number of them can be formed in 1 step. Let x be the number of parentheses in the set $\alpha$, and $y$ be the number of parentheses in the set $\beta$. Then,
\beq x+y= i_1+i_2+\ldots+i_m. \eeq
Consider a configuration $\kappa$ with all x of the parentheses from the set $\alpha$ and $0\leq z\leq y-1$ of the parentheses from the set $\beta$. Then, by the inductive hypothesis, $C_{\kappa}=(-1)^{x+z}$. Since one more step is required to reach the type $m_{(i_1,i_2,\ldots,i_m)}$ configuration in question, the contribution of $\kappa$ to $C_{m_{(i_1,i_2,\ldots,i_m)}}$ is $C_{\kappa}(-1)=(-1)^{x+z+1}$. There are $\binom{y}{z}$ such configurations since we have to choose z out of y parentheses from the set $\beta$. Therefore,
\begin{eqnarray}
C_{m_{(i_1,i_2,\ldots,i_m)}} &=& \binom{y}{0}(-1)^{x+1}+\binom{y}{1}(-1)^{x+2}+\binom{y}{2}(-1)^{x+3}+\binom{y}{3}(-1)^{x+4}+\ldots+\binom{y}{y-1}(-1)^{x+y} \nn\\
&=& (-1)^x\left[-1+\binom{y}{1}-\binom{y}{2}+\binom{y}{3}-\ldots+(-1)^y\binom{y}{y-1}\right]\nn\\
&=&(-1)^x(-1)^y=(-1)^{x+y}= (-1)^{i_1+i_2+\ldots+i_m }.
\end{eqnarray}
Thus we have proven the claim and shown that $C_{m_{(i_1,i_2,\ldots,i_m)}} = (-1)^{i_1+i_2+\ldots+i_m } $.

We now have the following expression for the lattice sum of a disconnected diagram with components $D_1,D_2,\ldots D_n$.  
\beq LS[D_1 \  dc \  D_2  \ dc \ldots dc  \ D_n]= \sum\limits_\kappa (-1)^{i_1+i_2+\ldots+i_m }LS[w_1]LS[w_2]\ldots LS[w_p]. \lab{ndisc}\eeq
Here, the sum is over the configurations $\kappa$ that one can make from the components $D_1,D_2,\ldots D_n$. Each configuration $\kappa$ has $p$ outer parentheses, $i_1$ type 1 parentheses, $\ldots$ $i_m$ type $m$ parentheses. $LS[w_i]$ represents the lattice sum of the overlapping of components inside the $i^{th}$ outer parenthesis.

\subsection{Calculation of the thermodynamic potential\label{thermopotential}}
\subsubsection{Partition function as a sum over configurations}
We return to our calculation of $\frac{Z}{Z_0}$ via the diagrammatic rules presented in section \ref{diagramsz}. Let $z_D$ denote the contribution to  $\frac{Z}{Z_0}$ of a connected diagram D. Let $z_{(D_1\ldots D_p)}$ denote the contribution to  $\frac{Z}{Z_0}$ of a disconnected diagram with components $D_1,D_2,\ldots D_p$. Let us recall that such a disconnected diagram will have a certain multiplicity which we denote by $\eta_{(D_1\ldots D_p)}$. Suppose the components $D_1,D_2,\ldots D_p$ are of orders $n_1,n_2,\ldots n_p$ in $t$ respectively. Then, the whole diagram is of order $n$ in $t$, where $n=n_1+n_2+\ldots+n_p$. In addition, out of the p components $D_1,D_2,\ldots D_p$, let us suppose that $k\leq p$ are distinct, with degeneracies $g_1,g_2,\ldots,g_k$, where $g_1+g_2+\ldots+g_k=p$. Then, we find that
\beq \eta_{(D_1\ldots D_p)} = \frac{n!}{n_1!n_2!\ldots n_p!}\frac{1}{g_1!g_2!\ldots g_k!}.\eeq
This factor comes about because we must distribute $n$ labelled lines among $p$ components with $n_i$ lines going to component $D_i$, but exchanging all of the lines of two identical components does not give a new distribution of lines. Now, instead of drawing  $\eta_{(D_1\ldots D_p)}$ diagrams with different distributions of lines, we draw only one such diagram with contribution 
\beq \eta_{(D_1\ldots D_p)}z_{(D_1\ldots D_p)}= \frac{n!}{n_1!n_2!\ldots n_p!}\frac{1}{g_1!g_2!\ldots g_k!} \frac{\beta^n}{n!}z_{B(D_1\ldots D_p)}LS[D_1 \  dc \  D_2  \ dc \ldots dc \  D_p]. \eeq

Here, $z_{B(D_1\ldots D_p)}$ indicates that we have dropped the factors $\frac{\beta^n}{n!}$ and $LS[D_1 \  dc \  D_2  \ dc \ldots dc \  D_p] $ from $z_{(D_1\ldots D_p)}$. Now, let $l$ denote the number of loops and $x$ denote the number of full sites in the disconnected diagram with components $D_1,D_2,\ldots D_p$. Let $l_i$ denote the number of loops and $x_i$ denote the number of full sites in the component $D_i$. Then, we have that
\begin{eqnarray} 
t^nm^l(-1)^{x-l} &=& t^{n_1}m^{l_1}(-1)^{x_1-l_1}t^{n_2}m^{l_2}(-1)^{x_2-l_2}\ldots t^{n_p}m^{l_p}(-1)^{x_p-l_p}, \nn\\
z_{B(D_1\ldots D_p)} &=& z_{BD_1}z_{BD_2}\ldots z_{BD_p}. 
\end{eqnarray} 
Therefore,
\beq \eta_{(D_1\ldots D_p)}z_{(D_1\ldots D_p)}=\frac{1}{g_1!g_2!\ldots g_k!}z_{nLD_1}z_{nLD_2}\ldots z_{nLD_p}LS[D_1 \  dc \  D_2 \  dc \ldots dc \  D_p]. \eeq
Here, $z_{nLD}$ indicates that we have dropped the lattice sum from $z_D$ . The partition function can now be expressed as
\beq \frac{Z}{Z_0}=1+\sum\limits_{p=1}^\infty \sum\limits_{(D_1\ldots D_p)}\frac{1}{g_1!g_2!\ldots g_k!}z_{nLD_1}z_{nLD_2}\ldots z_{nLD_p}LS[D_1  \ dc \  D_2 \  dc \ldots dc \  D_p]. \eeq
Here, the sum over $(D_1\ldots D_p)$ includes only one term for each set of connected components since the multiplicity from the different distributions of lines has already been taken into account. Plugging in our expression for $LS[D_1  \ dc \  D_2 \  dc \ldots dc \  D_p]$ from \disp{ndisc}, we obtain
\beq \frac{Z}{Z_0}=1+\sum\limits_{p=1}^\infty \sum\limits_{(D_1\ldots D_p)}\frac{1}{g_1!g_2!\ldots g_k!}z_{nLD_1}z_{nLD_2}\ldots z_{nLD_p}\sum\limits_\kappa (-1)^{i_1+i_2+\ldots+i_m }LS[w_1]LS[w_2]\ldots LS[w_q]. \eeq 
Here, the sum over $\kappa$ runs over all configurations that arise from the components $(D_1 \ldots D_p)$. However, a given configuration $\kappa$ can only come from a unique set of components $(D_1\ldots D_p)$. Hence, we can sum over all possible $\kappa$ directly, where in each term of the sum, by specifying $\kappa$, we automatically specify the $(D_1\ldots D_p)$ that it came from.
\beq \frac{Z}{Z_0}=1+\sum\limits_\kappa \frac{1}{g_1!g_2!\ldots g_k!}z_{nLD_1}z_{nLD_2}\ldots z_{nLD_p}(-1)^{i_1+i_2+\ldots+i_m }LS[w_1]LS[w_2]\ldots LS[w_q]. \lab{z4} \eeq
\subsubsection{Proof of the linked cluster theorem with respect to generalized connected components}
For each configuration $\kappa$, we relabel the components $(D_1\ldots D_p)$ according to which ones are contained in $w_i$. Those in $w_1$ are now labeled $(D_{w_11}D_{w_12}\ldots D_{w_1r_1})$ \ldots those in $w_q$ are now labeled $(D_{w_q1}D_{w_q2}\ldots D_{w_qr_q})$, where $r_1$ is the number of components in $w_1$, etc. We also count the number of different types of parentheses in the $w_j$. We denote the highest type of parenthesis (which is the outer parenthesis) in a given $w_j$ by $m_{w_j}$. We denote the number of type 1 parentheses in $w_j$ by $i_1(w_j)$, \ldots  the number of type $m_{w_j}$ parentheses in $w_j$ by $i_{m_{w_j}}(w_j)$. Note that $i_{m_{w_j}}(w_j)=1$, unless $w_j=D_i$ (i.e. $m_{w_j}=0$ ), in which case $i_{m_{w_j}}(w_j)\equiv0$. Then, we have that
\beq i_1+i_2 + \ldots + i_m = i_1(w_1)+\ldots+i_{m_{w_1}}(w_1)+\ldots+i_1(w_q)+\ldots+i_{m_{w_q}}(w_q). \eeq
We also make the following definition. For a particular $w$,
\beq z_w \equiv z_{nLD_{w1}}z_{nLD_{w2}}\ldots z_{nLD_{wr_w}}(-1)^{i_1(w)+i_2(w)+\ldots+i_{m_w}(w)}LS[w]. \lab{zw}\eeq
\disp{z4} now becomes
\beq \frac{Z}{Z_0}=1+\sum\limits_\kappa \frac{1}{g_1!g_2!\ldots g_k!}z_{w_1}z_{w_2}\ldots z_{w_q}. \label{z5}  \eeq

In a given configuration $\kappa$, all of the p components that go into making $\kappa$ are given the distinct labels $D_1\ldots D_p$, regardless of whether they are identical or not. However, there are distinct configurations which would be identical if we gave identical components the same label. As an example, consider configurations with three components which are all identical, but given the labels $D_1,D_2,D_3$. Then, the configuration $((D_1\cap D_2)\cap D_3)$ is distinct from the configuration $((D_1\cap D_3)\cap D_2)$. However, if all 3 components had the same label, the 2 configurations would be the same. All such configurations clearly have equal contributions to $\frac{Z}{Z_0}$. We therefore only consider configurations $\kappa^*$, in which identical components are given identical labels.
\beq \sum\limits_\kappa Z(\kappa) =\sum\limits_{\kappa^*} H(\kappa^*)Z(\kappa^*). \eeq
Here, $H(\kappa^*)$ is equal to the number of configurations $\kappa$ which collapse into $\kappa^*$ once identical components are given the same label. 

Consider an arbitrary configuration $\kappa$. It consists of $w_1,w_2,\ldots w_q$. Transform it to a $\kappa^*$ by giving identical components the same label. Then $\kappa^*$ consists of $w_1^*,w_2^*,\ldots w_q^*$, where $w_j$ is transformed to $w_j^*$ in the process of transforming $\kappa$ to $\kappa^*$. After the transformation, there may be some degeneracy among the $w_j^*$. Suppose there are $u$ distinct $w^*$:$w_1^*,w_2^*,\ldots w_u^*$ with degeneracies $s_1,s_2,\ldots,s_u$ respectively. Then, 
\beq H(\kappa^*) = \frac{g_1!g_2!\ldots g_k!}{s_1!s_2!\ldots s_u![\gamma(w_1^*)]^{s_1}[\gamma(w_2^*)]^{s_2}\ldots[\gamma(w_u^*)]^{s_u}}, \label{H} \eeq
where $\gamma(w_j^*)$ is the symmetry factor of $w_j^*$. By this we mean that if we momentarily give distinct labels to the identical components in  $w_j^*$, it is the number of ways to permute labels amongst {\bf identical} components and return to the same labeling. For example, if $w_j^*=((D_1\cap D_1)\cap(D_1\cap D_1))$, then $\gamma(w_j^*)=8$. If $w_j^*=((D_1\cap D_1)\cap(D_2\cap D_2))$, then $\gamma(w_j^*)=4$.  Plugging the expression for $H(\kappa^*)$ into \disp{z5}, we obtain
\beq \frac{Z}{Z_0}=1+\sum\limits_{\kappa^*}\frac{[z_{w^*_1}]^{s_1}[z_{w^*_2}]^{s_2}\ldots [z_{w^*_u}]^{s_u} }{s_1!s_2!\ldots s_u![\gamma(w_1^*)]^{s_1}[\gamma(w_2^*)]^{s_2}\ldots[\gamma(w_u^*)]^{s_u}}.\lab{z6} \eeq 
An arbitrary $\kappa^*$ has an arbitrary number (from $0$ to $\infty$) of each of the different possible $w^*$, except there has to be at least 1 $w^*$ of some kind for $\kappa^*$ to exist. However, the term in which there are zero of all of the possible $w^*$ is 1 and is therefore accounted for in the expression for $\frac{Z}{Z_0}$. Therefore,
\beq \frac{Z}{Z_0} = \exp\left[\sum\limits_{w^*}\frac{z_{w^*}}{\gamma(w^*)}\right], \eeq 
\beq \ln\left(\frac{Z}{Z_0}\right) = \sum\limits_{w^*}\frac{z_{w^*}}{\gamma(w^*)}.\label{logz} \eeq
We see that the $w^*$ are the ``generalized components" alluded to in section \ref{linked-cluster}. 
\subsubsection{Diagrammatic rules for calculating the thermodynamic potential}
We now have the following set of rules for the $n^{th}$ order contribution to $\ln(\frac{Z}{Z_0})$:
\begin{itemize}
\item[1] Choose $p \ge 1$ connected diagrams (not necessarily distinct) whose orders add up to $n$. If $p=1$, the overall diagram is connected and evaluated according to the rules for $\frac{Z}{Z_0}$. 
\item[2] If $p>1$, the overall diagram is disconnected. The p connected diagrams are now components in this disconnected diagram.
\item[3] Denote the contribution of a connected diagram $D$ to $\frac{Z}{Z_0}$ by $z_D$. If one removes the lattice sum factor from $z_D$, denote this by $z_{nLD}$. Multiply together the factors $z_{nLD}$ from the individual components.
\item[4] Create all distinct $w^*$ from the components by forming $i_1(w^*)$ type 1 parentheses, \ldots , $i_{m_{w^*}}(w^*)=1$ type $m_{w^*}$ parentheses. There is only one outer parenthesis in each $w^*$ which encloses all $p$ components. The contribution of each $w^*$ is $\frac{(-1)^{i_1(w^*)+\ldots+i_{m_{w^*}}(w^*)}LS[w^*]}{\gamma(w^*)}$, where $\gamma(w^*)$ is the symmetry factor of $w^*$ as explained below \disp{H}, and $LS[w^*]$ is the lattice sum of the overlapping of components represented by $w^*$. Sum the contributions over all $w^*$.
\item[5] Multiply the factor from 4 by the factor from 3.       
\end{itemize}

In addition to proving the above rules, the derivation shows that $\ln\frac{Z}{Z_0}$ is indeed the term in $\frac{Z}{Z_0}$ proportional to $L$. From \disp{zw}, we see that $z_{w*}$ is proportional to $L$. Therefore, the term proportional to $L$ on the RHS of \disp{z6} corresponds to $s_i=1$ for some $i$ and $s_j=0$ for $j \neq i$. Comparing this with \disp{logz}, we see that it is equal to $\ln\frac{Z}{Z_0}$. Therefore, an alternative to the above set of rules for $\ln\frac{Z}{Z_0}$ is to use the rules for $\frac{Z}{Z_0}$ from section \ref{diagramsz}, but to keep only the term proportional to $L$ in the lattice sum of a disconnected diagram. 

\section{Expansion for time dependent correlation functions}
\subsection{Diagrams for the numerator of the Green's function \label{diagramsG}}
The Green's function is defined as
\beq G_{jj'\sigma}(\tau) = -\langle X_{j}^{0\sigma}(\tau) X_{j'}^{\sigma0}\rangle \label{G1}, \eeq
where
\beq \langle O \rangle = \frac{Tr(e^{-\beta H} O)}{Z},  \eeq
and the time dependence is given by the Heisenberg representation. Furthermore, $0\leq \tau \leq \beta$, where $\beta$ is the inverse temperature. From \disp{G1}, the Green's function is written as
\beq G_{jj'\sigma}(\tau) = -\frac{Tr(e^{-\beta H} X_j^{0\sigma}(\tau)X_{j'}^{\sigma0})}{Z}. \eeq
Plugging in \disp{TN}, 
\beq G_{jj'\sigma}(\tau) = -\frac{e^{\mu\tau}\langle e^{(\tau-\beta)\hat{T}}X_j^{0\sigma}e^{-\tau\hat{T}}X_{j'}^{\sigma0}\rangle_0}{\frac{Z}{Z_0}}. \eeq
Finally, expanding the exponentials, we obtain 
\beq \resizebox{1.15\hsize}{!}{$G_{jj'\sigma}(\tau)(\frac{Z}{Z_0}) = -e^{\mu\tau}\sum\limits_{a=0b=0}^\infty \frac{(\beta-\tau)^a}{a!} \frac{\tau^b}{b!} \sum\limits_{\substack{j_1j_1'\ldots j_nj_n'\\\sigma_1\ldots \sigma_n}} t_{j_1j_1'}\ldots t_{j_nj_n'}\langle X_{j_1'}^{\sigma_10}X_{j_1}^{0\sigma_1}\ldots X_{j_a'}^{\sigma_a0}X_{j_a}^{0\sigma_a}X_j^{0\sigma}X_{j_{a+1}'}^{\sigma_{a+1}0}X_{j_{a+1}}^{0\sigma_{a+1}}\ldots X_{j_n'}^{\sigma_n0}X_{j_n}^{0\sigma_n}X_{j'}^{\sigma0}\rangle_0$}, \label{G2} \eeq
where $n=a+b$. This leads to the following rules for calculating the $n^{th}$ order contribution to $G_{jj'\sigma}(\tau)(\frac{Z}{Z_0})$:
\begin{itemize}
\item[1] Choose $0\leq a\leq n$. Set $b = n-a$. Draw the diagram as you would for $\frac{Z}{Z_0}$, except for the following changes. Begin drawing the diagram with a line labeled by $\sigma_{n.6}$ going into an empty site labeled by $j'$. The line does not come out of any site, it only goes into $j'$. In addition to this, after drawing the line labeled by $\sigma_{a+1}$, and before drawing the line labeled by $\sigma_a$, draw a line labeled by $\sigma_{a.5}$ going out of a site labeled $j$. The line does not go into any site, it only comes out of $j$. $j$ and $j'$ may label the same site.
\item[2]Insert a factor of  $ \frac{-e^{\mu \tau}}{a!b!}(\beta-\tau)^a\tau^b$.
\item[3] Insert a factor of $\frac{\rho}{m}$ for each filled circle, and $(1-\rho)$ for each empty circle. 
\item[4] Compute the multiplicity of the diagram, keeping in mind that $j$ and $j'$ are fixed sites on the lattice. Insert a factor of $t^n$.
\item[5] Pair the lines at each site in the same way as for $\frac{Z}{Z_0}$, including the $\sigma_{n.6}$ and $\sigma_{a.5}$ lines. Now, find the line which is paired with the $\sigma_{n.6}$ line on the site $j'$. Follow the same process as for $\frac{Z}{Z_0}$ until you reach the  $\sigma_{a.5}$ line. This completes the loop started by the $\sigma_{n.6}$ line. Find the line with the highest remaining spin label, and continue to the break the diagram into loops just as for $\frac{Z}{Z_0}$. Let $l$ be the number of loops. Let $x$ be the number of full sites in the diagram. Insert a factor of $(-1)^{x+1-l}m^{l-1}$.
\end{itemize}
\subsubsection{Proof of the rules for the numerator of the Green's function}
The rules for calculating $G_{jj'\sigma}(\tau)(\frac{Z}{Z_0})$ are modified from those for calculating $\frac{Z}{Z_0}$ by introducing the external line $\sigma_{n.6}$ on the site $j'$ to account for $X_{j'}^{\sigma0}$, and the external line $\sigma_{a.5}$ on the site $j$ to account for $X_j^{0\sigma}$. We now state the proof of rule 5 for $G_{jj'\sigma}(\tau)(\frac{Z}{Z_0})$, which is similar to the proof of rule 5 for $\frac{Z}{Z_0}$, given in section \ref{proofspinsign}. 

The order of the numbers in the expectation value in \disp{G2} is $1'12'2\ldots a'a a_{.5}(a+1)'(a+1)\ldots n'n n_{.6}'$. This is equivalent to the order $1'12'2\ldots a'a (a+1)'(a+1)\ldots n'n a_{.5} n_{.6}'$, since $a_{.5}$ has to be moved past a certain number of pairs to get to the left of $n_{.6}'$. The first loop is started by the pair $(x, n_{.6}')$ on an empty site. This corresponds to the $(x_{p-1},x_p')$ pair on this site. The next pair in this loop is either $(y,x')$ or $(x',y)$, where the latter can only correspond to the $(x_1',x_p)$ pair on a full site. The loop now becomes $\pm (x'xyn_{.6}')$ where $(y,x')$ gives $+$ and $(x',y)$ gives $-$. This process continues until we come across $a_{.5}$, at which point we complete the first loop.

The subsequent loops are made in the same way as for $\frac{Z}{Z_0}$. We make the same arguments  to show that each subsequent loop is started by the $(x_1',x_p)$ pair on a full site. However, in the case of $G_{jj'\sigma}(\tau)(\frac{Z}{Z_0})$, this does not apply to the first loop. Therefore, only $l-1$ full sites don't contribute a minus sign, and the sign of the diagram is $(-1)^{x-(l-1)}=(-1)^{x-l+1}$. 

For the spin sum we note that  $\sigma_{n.6}=\sigma_{a.5}=\sigma$. Thus, the spin of the first loop is fixed to be $\sigma$. Hence, this loop does not give a factor of m. Therefore, the spin sum is $m^{l-1}$. The overall factor from the sign and spin sum is $(-1)^{x-l+1}m^{l-1}$ as stated in rule 5.

\subsection{ Calculation of the Green's function\label{calcGreen}}
\subsubsection{Numerator of the Green's function as a sum over configurations \label{numGreen}}
To proceed further with our calculation of the Green's function, we need to address the issue of disconnected diagrams in $G_{jj'\sigma}(\tau)(\frac{Z}{Z_0})$. In a disconnected diagram, there will be one component, which we denote by $D_G$, which will contain the external lines. The other components will be the same as those in the diagrams for $\frac{Z}{Z_0}$. Let $G_{D_G}$ be the contribution of $D_G$ as calculated by the rules for $G_{jj'\sigma}(\tau)(\frac{Z}{Z_0})$. Let $G_{(D_GD_1\ldots D_p)}$ be the contribution of a disconnected diagram with components $D_G D_1\ldots D_p$ as calculated by these rules. Consider a disconnected diagram of order $n$, with $0\leq a\leq n$, comprised of components $D_GD_1\ldots D_p$, where $D_G$ is of order $c$, and $D_1\ldots D_p$ are of orders $n_1\ldots n_p$ respectively. Let $f\equiv n_1+n_2+\ldots+n_p$. Then, $f=n-c$. In general, $D_G$ has $\alpha$ lines numbered lower than $\sigma_{a.5}$, and $\delta\equiv c-\alpha$ lines numbered higher than $\sigma_{a.5} $(excluding $\sigma_{n.6}$). Additionally, there could be some degeneracy among the components $D_1\ldots D_p$. Assume that out of these, $k$ are distinct, with degeneracy $g_1 \ldots g_k$. Then, the number of different ways to distribute lines among the components $D_GD_1\ldots D_p$ is:
\beq \eta_{a(D_GD_1\ldots D_p)} = \binom{a}{\alpha}\binom{b}{\delta}\frac{f!}{n_1!\ldots n_p!}\frac{1}{g_1!\ldots g_k!}.\label{etaGa}\eeq
Hence, for a given choice of $a$, the components $D_GD_1\ldots D_p$ make the following contribution to $G_{jj'\sigma}(\tau)(\frac{Z}{Z_0})$:
\beq \eta_{a(D_GD_1\ldots D_p)}G_{(D_GD_1\ldots D_p)} = \binom{a}{\alpha}\binom{b}{\delta}\frac{f!}{n_1!\ldots n_p!}\frac{1}{g_1!\ldots g_k!}\frac{(-e^{\mu\tau})}{a!b!}(\beta-\tau)^a\tau^bG_{B(D_GD_1\ldots D_p)}LS[D_G \  dc \  D_1  \ dc \ldots dc \  D_p], \eeq
where $G_{B(D_GD_1\ldots D_p)}$ is $G_{(D_GD_1\ldots D_p)}$ without the factors $\frac{(-e^{\mu\tau})}{a!b!}(\beta-\tau)^a\tau^b$ and  $LS[D_G  \ dc  \ D_1 \ dc \ldots dc  \ D_p] $. However, there are multiple values of $0\leq a\leq n$ for which a disconnected diagram can have the components $D_GD_1\ldots D_p$. The values of $a$ are restricted by the values of $\alpha$ and $\delta$ for the given $D_G$. In particular, $\alpha \leq a$ and $\delta \leq n-a$. Therefore, we have
\beq \alpha\leq a\leq n- \delta. \eeq
Hence, the total contribution of the components $D_GD_1\ldots D_p$ to $G_{jj'\sigma}(\tau)(\frac{Z}{Z_0})$ is:
\barray \eta_{(D_GD_1\ldots D_p)}G_{(D_GD_1\ldots D_p)} = \sum\limits_{a=\alpha}^{n-\delta}&&\frac{1}{\alpha!(a-\alpha)!}\frac{1}{\delta!(b-\delta)!}\frac{f!}{n_1!\ldots n_p!}\frac{1}{g_1!\ldots g_k!}\nn\\
&& \times (-e^{\mu\tau})(\beta-\tau)^{a-\alpha}\tau^{b-\delta}(\beta-\tau)^\alpha\tau^\delta G_{B(D_GD_1\ldots D_p)}LS[D_G  \ dc \  D_1 dc \ldots dc \ D_p]. \nn\\
\label{G3} \earray
We also have the following equalities:
\beq G_{B(D_GD_1\ldots D_p)} = G_{BD_G}z_{B(D_1\ldots D_p)}, \eeq
\beq G_{nLD_G} = \frac{(-e^{\mu\tau})}{\alpha!\delta!}(\beta-\tau)^\alpha \tau^\delta G_{BD_G}. \eeq
Furthermore, using the definitions of $b$, $\delta$, and $f$, one can show that
\beq \sum\limits_{a=\alpha}^{n-\delta}\frac{1}{(a-\alpha)!}\frac{1}{(b-\delta)!}(\beta-\tau)^{a-\alpha}\tau^{b-\delta}=\sum\limits_{a=0}^{f}\frac{1}{a!}\frac{1}{(f-a)!}(\beta-\tau)^a\tau^{f-a}= \frac{\beta^f}{f!}. \eeq
Hence, \disp{G3} simplifies to:
\beq \eta_{(D_GD_1\ldots D_p)}G_{(D_GD_1\ldots D_p)} = \frac{1}{g_1!\ldots g_k!}G_{nLD_G}z_{nLD_1}\ldots z_{nLD_p}LS[D_G  \ dc \ D_1 \  dc \ldots dc \  D_p]. \eeq
Therefore,
\beq G_{jj'\sigma}(\tau)\left(\frac{Z}{Z_0}\right)=\sum\limits_{D_G}\sum\limits_{p=0}^\infty \sum\limits_{(D_1\ldots D_p)}\frac{1}{g_1!\ldots g_k!}G_{nLD_G}z_{nLD_1}\ldots z_{nLD_p}LS[D_G \  dc \ D_1 \ dc \ldots dc \ D_p]. \eeq
$LS[D_G  \ dc \  D_1 \  dc \ldots dc \ D_p] $ is evaluated in terms of configurations in the same way as $LS[D_1 \ dc \ldots dc \ D_p]$ from $\frac{Z}{Z_0}$ was, except that now in each $\kappa$, there will be one $w$ which contains $D_G$. We shall denote it by $w_G$. The other $w$ will be the same as those found in the calculation of $\frac{Z}{Z_0}$. Thus, we obtain 
\beq G_{jj'\sigma}(\tau)\left(\frac{Z}{Z_0}\right)=\sum\limits_{D_G}\sum\limits_{p=0}^\infty \sum\limits_{(D_1\ldots D_p)}\frac{1}{g_1!\ldots g_k!}G_{nLD_G}z_{nLD_1}\ldots z_{nLD_p}\sum\limits_\kappa(-1)^{i_1+\ldots + i_m}LS[w_G]LS[w_1]\ldots LS[w_q]. \eeq
A given $\kappa$ can only come from a unique $(D_G D_1\ldots D_p)$. Therefore,
\beq G_{jj'\sigma}(\tau)\left(\frac{Z}{Z_0}\right)=\sum\limits_\kappa \frac{1}{g_1!\ldots g_k!}G_{nLD_G}z_{nLD_1}\ldots z_{nLD_p}(-1)^{i_1+\ldots + i_m}LS[w_G]LS[w_1]\ldots LS[w_q]. \label{G4} \eeq
\subsubsection{Cancellation of the denominator of the Green's function through the linked cluster theorem}
We define
\beq G_{w_G}\equiv G_{nLD_G(w_G)} z_{nLD_{w_G1}}z_{nLD_{w_G2}}\ldots z_{nLD_{w_Gr_{w_G}}}(-1)^{i_1(w_G)+i_2(w_G)+\ldots+i_{m_{w_G}}(w_G)}LS[w_G]. \eeq
Then, \disp{G4} simplifies to
\beq G_{jj'\sigma}(\tau)\left(\frac{Z}{Z_0}\right)=\sum\limits_\kappa \frac{1}{g_1!\ldots g_k!} G_{w_G}z_{w_1}\ldots z_{w_q}. \label{G5}  \eeq
Just as in the case of $\frac{Z}{Z_0}$, we perform the transformation $\kappa \rightarrow \kappa^*$ by giving identical components identical labels. In the process, $w_G\rightarrow w^*_G$ and  $w_j\rightarrow w^*_j$. After the transformation, there may be some degeneracy in the $w_1^*\ldots w_q^*$. Suppose there are $u$ distinct $w_j^*$: $w_1^*\ldots w_u^*$ with degeneracies $s_1 \ldots s_u$ respectively. There can only be one $w^*_G$ because there is only one $D_G$ in any diagram. Therefore, the multiplicity factor for the number of configurations $\kappa$ which correspond to a single $\kappa^*$ is
\beq H(\kappa^*) = \frac{g_1!g_2!\ldots g_k!}{s_1!s_2!\ldots s_u![\gamma(w_1^*)]^{s_1}[\gamma(w_2^*)]^{s_2}\ldots[\gamma(w_u^*)]^{s_u}\gamma(w_G^*)}, \eeq
and \disp{G5} becomes
\beq G_{jj'\sigma}(\tau)\left(\frac{Z}{Z_0}\right)=\sum\limits_{\kappa^*}\frac{G_{w_G^*}[z_{w^*_1}]^{s_1}[z_{w^*_2}]^{s_2}\ldots [z_{w^*_u}]^{s_u} }{s_1!s_2!\ldots s_u![\gamma(w_1^*)]^{s_1}[\gamma(w_2^*)]^{s_2}\ldots[\gamma(w_u^*)]^{s_u}\gamma(w_G^*)}. \label{G6} \eeq 
 An arbitrary $\kappa^*$ has one of the $w^*_G$ and an arbitrary number from $0$ to $\infty$ of each of the $w^*$. Therefore,
 \beq G_{jj'\sigma}(\tau)\left(\frac{Z}{Z_0}\right)=\sum\limits_{w_G^*}\frac{G_{w_G^*}}{\gamma(w_G^*)}\exp\left[\sum\limits_{w^*}\frac{z_{w^*}}{\gamma(w^*)}\right]. \eeq
 Recalling that $\frac{Z}{Z_0} = \exp[\sum\limits_{w^*}\frac{z_{w^*}}{\gamma(w^*)}]$, we have that
 \beq G_{jj'\sigma}(\tau)=\sum\limits_{w_G^*}\frac{G_{w_G^*}}{\gamma(w_G^*)}.\label{G7}\eeq
 \subsubsection{Diagrammatic rules for the calculation of the Green's function}
 This leads to the following set of rules for the $n^{th}$ order contribution to $G_{jj'\sigma}(\tau)$:
 \begin{itemize}
\item[1]   Choose one connected diagram from $G_{jj'\sigma}(\tau)(\frac{Z}{Z_0})$ and $p \geq 0$ connected diagrams (not necessarily distinct) from $\frac{Z}{Z_0}$. The orders of the $p+1$ diagrams must add up to $n$. If $p=0$, then the overall diagram is connected and must be evaluated according to the rules for $G_{jj'\sigma}(\tau)(\frac{Z}{Z_0})$.
\item[2] If $p>0$, the overall diagram is disconnected. The $p+1$ connected diagrams are now components in this disconnected diagram.
\item[3] Denote the contribution of a connected diagram $D_G$ to $G_{jj'\sigma}(\tau)(\frac{Z}{Z_0})$ by $G_{D_G}$. If one removes the lattice sum factor from $G_{D_G}$, denote this by $G_{nLD_G}$. Multiply $G_{nLD_G}$ from the component $D_G$ together with $z_{nLD_1} \ldots z_{nLD_p}$ from the $p$ components $D_1 \ldots D_p$. 
\item[4] Create all distinct $w^*$ from the components by forming $i_1(w^*)$ type 1 parentheses, \ldots , $i_{m_{w^*}}(w^*)=1$ type $m_{w^*}$ parentheses. There is only one outer parenthesis in each $w^*$ which encloses all $p+1$ components. The contribution of each $w^*$ is $\frac{(-1)^{i_1(w^*)+\ldots+i_{m_{w^*}}(w^*)}LS[w^*]}{\gamma(w^*)}$, where $\gamma(w^*)$ is the symmetry factor of $w^*$, and $LS[w^*]$ is the lattice sum of the overlapping of components represented by $w^*$. Sum the contributions over all $w^*$.
\item[5] Multiply the factor from 4 by the factor from 3.   
\end{itemize}

In addition to proving the above rules, the derivation shows that $G_{jj'\sigma}(\tau)$ is indeed the term in $G_{jj'\sigma}(\tau)(\frac{Z}{Z_0})$ independent to $L$. From \disp{zw}, we see that $z_{w*}$ is proportional to $L$. Therefore, the term independent $L$ on the RHS of \disp{G6} corresponds to $s_i=0$ for all $i$. Comparing this with \disp{G7}, we see that it is equal to $G_{jj'\sigma}(\tau)$. Therefore, an alternative to the above set of rules for $G_{jj'\sigma}(\tau) $ is to use the rules for $G_{jj'\sigma}(\tau)(\frac{Z}{Z_0})$ from section \ref{diagramsG}, but to keep only the term independent of $L$ in the lattice sum of a disconnected diagram.
\subsection{Examples and results for the Green's function \label{Greenfour}}
\subsubsection{Examples from $0^{th}$ to $2^{nd}$ order} 
We now show some examples of diagrams for $G_{jj'\sigma}(\tau) $, with the contribution written next to the diagram. In zeroth order, there is only one diagram:
\begin{center}
\fcolorbox{white}{white}{
  \begin{picture}(128,102) (51,-27)
    \SetWidth{1.0}
    \SetColor{Black}
    \COval(96,22)(16,16)(0){Black}{White}
    \Line[arrow,arrowpos=0.5,arrowlength=5,arrowwidth=2,arrowinset=0.2](96,70)(96,38)
    \Line[arrow,arrowpos=0.5,arrowlength=5,arrowwidth=2,arrowinset=0.2](96,6)(96,-26)
    \Text(96,54)[lb]{\Large{\Black{$\sigma_{0.6}$}}}
    \Text(96,-10)[lb]{\Large{\Black{$\sigma_{0.5}$}}}
    \Text(35,22)[lb]{\Large{\Black{$a=0$}}}
    \Text(144,22)[lb]{\Large{\Black{$-e^{\mu\tau}(1-\rho)\delta_{jj'}$}}}
    \Text(84,18)[lb]{{\Black{$j'=j$}}}
  \end{picture}
}
\end{center}
In first order, there are 2 diagrams:
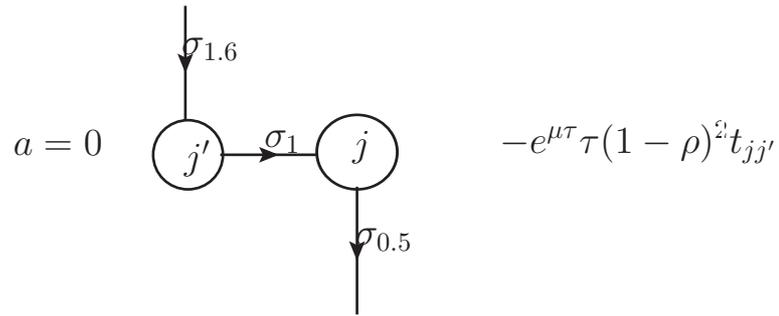
\begin{center}
\fcolorbox{white}{white}{
  \begin{picture}(256,118) (51,-23)
    \SetWidth{1.0}
    \SetColor{Black}
    \COval(113,38)(13,13)(0){Black}{White}
    \COval(176,39)(14,15)(0){Black}{White}
    \Line[arrow,arrowpos=0.5,arrowlength=5,arrowwidth=2,arrowinset=0.2](112,94)(112,51)
    \Line[arrow,arrowpos=0.5,arrowlength=5,arrowwidth=2,arrowinset=0.2](125,38)(161,38)
    \Line[arrow,arrowpos=0.5,arrowlength=5,arrowwidth=2,arrowinset=0.2](176,26)(176,-22)
    \Text(111,74)[lb]{\Large{\Black{$\sigma_{1.6}$}}}
    \Text(142,39)[lb]{\Large{\Black{$\sigma_1$}}}
    \Text(176,2)[lb]{\Large{\Black{$\sigma_{0.5}$}}}
    \Text(48,38)[lb]{\Large{\Black{$a=0$}}}
    \Text(112,30)[lb]{\Large{\Black{$j'$}}}
    \Text(175,35)[lb]{\Large{\Black{$j$}}}
    \Text(230,35)[lb]{\Large{\Black{$-e^{\mu\tau}\tau(1-\rho)^2t_{jj'}$}}}
  \end{picture}
}
\end{center}
\begin{center}
\fcolorbox{white}{white}
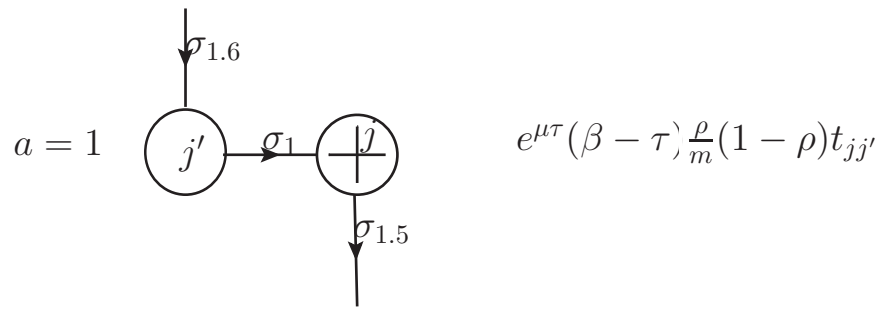
{
  \begin{picture}(240,115) (35,-7)
    \SetWidth{1.0}
    \SetColor{Black}
    \COval(96,52)(16,15)(0){Black}{White}
    \COval(160,51)(15,15)(90.0){Black}{White}\Line(149.393,51)(170.607,51)\Line(160,61.607)(160,40.393)
    \Line[arrow,arrowpos=0.5,arrowlength=5,arrowwidth=2,arrowinset=0.2](96,106)(96,69)
    \Line[arrow,arrowpos=0.5,arrowlength=5,arrowwidth=2,arrowinset=0.2](110,51)(145,51)
    \Line[arrow,arrowpos=0.5,arrowlength=5,arrowwidth=2,arrowinset=0.2](159,36)(160,-6)
    \Text(96,87)[lb]{\Large{\Black{$\sigma_{1.6}$}}}
    \Text(125,51)[lb]{\Large{\Black{$\sigma_1$}}}
    \Text(159,18)[lb]{\Large{\Black{$\sigma_{1.5}$}}}
    \Text(32,51)[lb]{\Large{\Black{$a=1$}}}
    \Text(94,45)[lb]{\Large{\Black{$j'$}}}
    \Text(162,52)[lb]{\Large{\Black{$j$}}}
    \Text(220,48)[lb]{\Large{\Black{$e^{\mu\tau}(\beta-\tau)\frac{\rho}{m}(1-\rho)t_{jj'}$}}}
  \end{picture}
}
\end{center}
In second order, one can have a disconnected diagram:
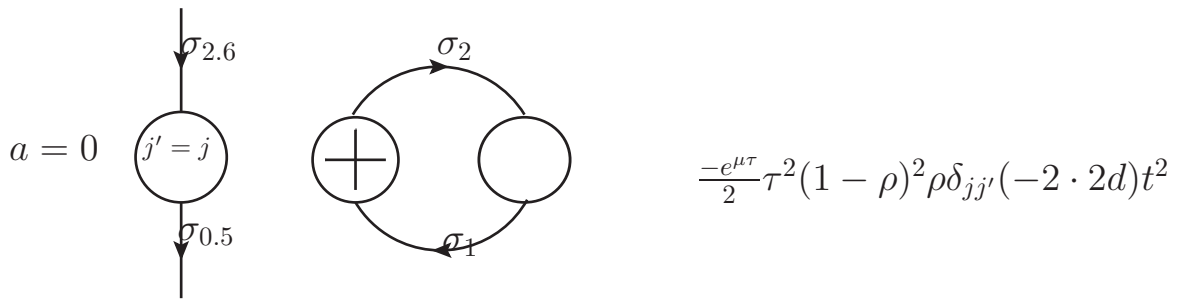
\begin{center}
\fcolorbox{white}{white}{
  \begin{picture}(436,150) (0,0)
    \SetWidth{1.0}
    \SetColor{Black}
    \COval(96,54)(17,17)(0){Black}{White}
    \COval(161,53)(16,16)(90.0){Black}{White}\Line(149.686,53)(172.314,53)\Line(161,64.314)(161,41.686)
    \COval(224,53)(16,17)(0){Black}{White}
    \Line[arrow,arrowpos=0.5,arrowlength=5,arrowwidth=2,arrowinset=0.2](96,110)(96,71)
    \Line[arrow,arrowpos=0.5,arrowlength=5,arrowwidth=2,arrowinset=0.2](96,37)(96,1)
    \Arc[arrow,arrowpos=0.5,arrowlength=5,arrowwidth=2,arrowinset=0.2,clock](192,50.556)(37.444,148.716,31.284)

    \Arc[arrow,arrowpos=0.5,arrowlength=5,arrowwidth=2,arrowinset=0.2,clock](192.713,55.89)(36.912,-28.99,-149.22)
    \Text(96,91)[lb]{\Large{\Black{$\sigma_{2.6}$}}}
    \Text(95,21)[lb]{\Large{\Black{$\sigma_{0.5}$}}}
    \Text(192,91)[lb]{\Large{\Black{$\sigma_2$}}}
    \Text(194,17)[lb]{\Large{\Black{$\sigma_1$}}}
    \Text(82,53)[lb]{{\Black{$j'=j$}}}
    \Text(32,53)[lb]{\Large{\Black{$a=0$}}}
    \Text(289,37)[lb]{\Large{\Black{$\frac{-e^{\mu\tau}}{2}\tau^2(1-\rho)^2\rho\delta_{jj'}(-2\cdot2d)t^2$}}}
  \end{picture}
}
\end{center}
\subsubsection{Green's function to fourth order in $\beta t$ with $m$ spin species on a d-dimensional hypercube}
We have used the above rules to calculate the Green's function to 4th order in t on a hypercube in $d$ dimensions with $m$ spin species:
\begin{eqnarray}
G_\sigma^{(0)}(\vk,\omega_k)&=&\frac{m (-\rho )+m+\rho }{m z}, \nonumber\\
G_\sigma^{(1)}(\vk,\omega_k)&=& \frac{\epsilon _k (m (-\rho )+m+\rho )^2}{m^2 z^2}, \nonumber\\
G_\sigma^{(2)}(\vk,\omega_k)&=&\frac{2 d (m-1) t^2 \rho  (m (\rho -2)-\rho ) (m (\rho -1)-\rho )}{m^3
   z^3}-\frac{2 d (m-1) t^2 \beta  (\rho -1) \rho }{m z^2}\nonumber\\
   &&-\frac{d (m-1)
   t^2 \beta ^2 (\rho -1) \rho  (2 \rho -1)}{m z}
   +\frac{\epsilon _k^2 (m
   (-\rho )+m+\rho )^3}{m^3 z^3},\nonumber\\
G_\sigma^{(3)}(\vk,\omega_k)&=&-\frac{(4 d-1) (m-1) t^2 \rho  \epsilon _k (m (\rho -2)-\rho ) (m (-\rho
   )+m+\rho )^2}{m^4 z^4}\nonumber\\
   &&+\frac{2 (2 d-1) (m-1) t^2 \beta  (\rho -1) \rho
    \epsilon _k (m (\rho -1)-\rho )}{m^2 z^3}\nonumber\\
    &&+\frac{(m-1) t^2 \beta ^2
   (\rho -1) \rho  \epsilon _k ((d (4 \rho -2)-\rho ) (m (\rho -1)-\rho
   )-\rho )}{m^2 z^2}+\frac{\epsilon _k^3 (m (-\rho )+m+\rho )^4}{m^4
   z^4},\nonumber\\
G_\sigma^{(4)}(\vk,\omega_k)&=&-\frac{2 (3 d-1) (m-1) t^2 \rho  \epsilon _k^2 (m (\rho -2)-\rho ) (m
   (-\rho )+m+\rho )^3}{m^5 z^5}\nonumber\\
   &&-\frac{2 (3 d-2) (m-1) t^2 \beta  (\rho
   -1) \rho  \epsilon _k^2 (m (-\rho )+m+\rho )^2}{m^3 z^4}\nonumber\\
   &&-\frac{(m-1)
   t^2 \beta ^2 (\rho -1) \rho  \epsilon _k^2 (m (\rho -1)-\rho ) ((d (6
   \rho -3)-2 \rho ) (m (\rho -1)-\rho )-2 \rho )}{m^3 z^3}\nonumber\\
   &&+\frac{2 d
   (m-1) t^4 \beta  (\rho -1) \rho  \left(-4 d m^2+2 (3 d-2) (m-1)^2 \rho
   ^2+m \rho  (-6 d (m-3)+5 m-11)\right)}{m^3 z^4}\nonumber\\
   &&+{\textstyle\frac{d (m-1) t^4
   \beta ^2 (\rho -1) \rho  \left(-4 d m^2+4 (3 d-1) (m-1)^2 \rho ^3-2
   (m-1) \rho ^2 (3 d (5 m-1)-4 m+2)+m \rho  (2 d (9 m-7)-3
   m+5)\right)}{m^3 z^3}}\nonumber\\
   &&-\frac{d (m-1) t^4 \beta ^3 (\rho -1) \rho 
   \left(m^2 \left(d \left(26 \rho ^2-28 \rho +6\right)+2 (5-4 \rho )
   \rho -3\right)+8 (d-1) m (\rho -1) \rho +2 (d-1) \rho ^2\right)}{3 m^3
   z^2}\nonumber\\
   &&-{\textstyle \frac{d (m-1) t^4 \beta ^4 (\rho -1) \rho  \left(m^2 (2 d (\rho 
   (\rho  (52 \rho -81)+34)-3)-2 \rho  (\rho  (16 \rho -27)+13)+3)+16
   (d-1) m (\rho -1) \rho  (2 \rho -1)+2 (d-1) \rho ^2 (4 \rho
   -3)\right)}{12 m^3 z}}\nonumber\\
   &&+{\textstyle \frac{2 d (m-1) t^4 \rho  (m (\rho -1)-\rho )
   \left(10 (1-2 d) m^3+2 m^2 \rho  (d (8 m-2)-6 m+3)-(m-1)^3 \rho ^3+4
   (m-1)^2 m \rho ^2\right)}{m^5 z^5}}\nonumber\\
   &&+\frac{\epsilon _k^4 (m (-\rho
   )+m+\rho )^5}{m^5 z^5}.   
\end{eqnarray}
\beq z\equiv i\omega_k + \mu. \eeq
In the case where $m=1$, this should reduce to the answer for a single species free Fermi gas. We can see that this is indeed the case as all the terms but the free propagator vanish. When we set $m=2$ and $d=2$, we recover the expressions obtained from the Metzner expansion \cite{EhsanMetzner}. Using the formula
\beq n-1 = G_{jj\sigma}(\tau=0),\eeq
we can convert to expressions where n is the independent parameter and hence t-independent.

\begin{eqnarray}
&&G_\sigma^{(0)}(\vk,\omega_k)=\frac{m (-n)+m+n}{m z}, \nonumber\\
&&G_\sigma^{(1)}(\vk,\omega_k)= \frac{\epsilon _k (m (-n)+m+n)^2}{m^2 z^2},\nonumber\\
&& G_\sigma^{(2)}(\vk,\omega_k)=\frac{2 d (m-1) n t^2 (m (n-2)-n) (m (n-1)-n)}{m^3 z^3}-\frac{d t^2 \beta
    (m (n-1)+n)}{m z^2}+\frac{\epsilon _k^2 (m (-n)+m+n)^3}{m^3 z^3},\nonumber\\
&& G_\sigma^{(3)}(\vk,\omega_k)=-\frac{(4 d-1) (m-1) n t^2 \epsilon _k (m (n-2)-n) (m (-n)+m+n)^2}{m^4
   z^4}\nonumber\\
   &&-\frac{2 t^2 \beta  \epsilon _k (m (n-1)-n) (d (m-(m+1) n)+(m-1)
   (n-1) n)}{m^2 z^3}+\frac{\epsilon _k^3 (m (-n)+m+n)^4}{m^4
   z^4}-\frac{(m-1)^2 (n-1)^2 n^2 t^2 \beta ^2 \epsilon _k}{m^2 z^2},\nonumber\\
&&G_\sigma^{(4)}(\vk,\omega_k)=\frac{2 (3 d-1) (m-1) n t^2 \epsilon _k^2 (m (n-2)-n) (m (n-1)-n)^3}{m^5
   z^5}\nonumber\\
   &&+\frac{t^2 \beta  \epsilon _k^2 (m (-n)+m+n)^2 (4 (m-1) (n-1) n-3
   d (m (n-1)+n))}{m^3 z^4}\nonumber\\
   &&+\frac{2 d (m-1) n t^4 \beta  \left(d
   \left(m^2 (n-1) (9 n-10)+3 m n (2 n-3)-3 n^2\right)+(1-n) n \left(4
   (m-1)^2 n+(11-5 m) m\right)\right)}{m^3 z^4}\nonumber\\
   &&+{\textstyle \frac{d t^4 \beta ^2
   \left(d m \left(m^2+2 (m-1)^2 n^3-2 ((m-4) m+1) n^2-m (m+3)
   n\right)+(1-m) (n-1) n^2 \left(4 (m-1)^2 n^2-4 (m-1) (2 m-1) n+m (3
   m-5)\right)\right)}{m^3 z^3}}\nonumber\\
   &&+{\textstyle \frac{d t^4 \beta ^3 \left(-2 (m+1) n^3
   \left((d-7) m^2+10 (d-1) m+d-1\right)+m^2 n (-8 d (m+2)+17 m+25)+2 m
   n^2 (d (5 m (m+4)+11)+(-2 m-1) (7 m+11))-3 m^3\right)}{12 m^3
   z^2}}\nonumber\\
   &&+\frac{2 d (m-1) n t^4 (m (n-1)-n) \left(10 (1-2 d) m^3+2 m^2 n (d
   (8 m-2)-6 m+3)-(m-1)^3 n^3+4 (m-1)^2 m n^2\right)}{m^5
   z^5}\nonumber\\
   &&+\frac{\epsilon _k^4 (m (-n)+m+n)^5}{m^5 z^5}+\frac{2 (m-1)^2
   (n-1)^2 n^2 t^2 \beta ^2 \epsilon _k^2 (m (n-1)-n)}{m^3 z^3}.  \label{G4n}
\end{eqnarray}
\beq z\equiv i\omega_k + \mu^{(0)}.\eeq
\beq \rho(\mu^{(0)})\equiv n.\eeq

\subsection{The infinite spin species limit. \label{infinitespins}}
It is interesting to see how the above expressions simplify in the limit of infinite spin species, i.e. as $m \to \infty$. In this case, the Green's function may be written in the form
\beq  G(k) = \frac{1-n}{z-(1-n)\epsilon_k-\Sigma_{DM}(k)},\eeq 
where $\Sigma_{DM}(k)$ is the Dyson-Mori self energy \cite{Anatomy}, which has a finite value as $i\omega_k\to\infty$. The high frequency limit of the Green's function is therefore $\lim_{i\omega_k\to\infty}G(k)=\frac{1-n}{i\omega_k}$, as can be seen explicitly from the anti-commutation relations of the Hubbard X operators:
\beq \langle\{X_i^{0\sigma},X_j^{\sigma0}\}\rangle =\delta_{ij}\langle X_i^{00}+X_i^{\sigma\sigma}\rangle= \delta_{ij}\langle 1-\sum_{\sigma'\neq\sigma}X_i^{\sigma'\sigma'}\rangle=\delta_{ij}[1-(m-1)\frac{n}{m}].\eeq
Taking the $m \to \infty$ limit of the above equation gives the high frequency coefficient $1-n$, while $m=2$ gives the usual coefficient of $1-\frac{n}{2}$ \cite{ECQL}.
Using \disp{G4n}, we derive a high temperature expansion for $\Sigma_{DM}(k)$ in the $m\to\infty$ limit:
\begin{eqnarray}
&&\Sigma_{DM}^{(0)}(\vk,\omega_k)=0, \nonumber\\
&&\Sigma_{DM}^{(1)}(\vk,\omega_k)= 0,\nonumber\\
&& \Sigma_{DM}^{(2)}(\vk,\omega_k)=d t^2 \beta -\frac{2 d (n-2) n t^2}{z},\nonumber\\
&& \Sigma_{DM}^{(3)}(\vk,\omega_k)=(n-1) n^2 t^2 \beta ^2 \epsilon _k-\frac{(n-2) (n-1) n t^2 \epsilon
   _k}{z^2}+\frac{2 (n-1) n t^2 \beta  \epsilon _k}{z},\nonumber\\
&&\Sigma_{DM}^{(4)}(\vk,\omega_k)=\frac{d n^2 t^4 \beta ^2 (-2 d+4 (n-2) n+3)}{z}-\frac{2 d (2 d-1) n (n
   ((n-4) n+12)-10) t^4}{z^3}\nonumber\\
&&-\frac{2 d n t^4 \beta  (d (7 n-6)+(5-4 n)
   n)}{z^2}+\frac{1}{12} d t^4 \beta ^3 (2 n (d (n-4)-7 n+7)-3).  \label{SigDM4n}
\end{eqnarray}
The connection between this high temperature expansion for the Dyson-Mori self-energy in the $m\to\infty$ limit and slave boson techniques \cite{ReadNewns,Coleman,KotliarLiu} is an interesting direction for further study.

\subsection{Calculation of time-dependent density-density and spin-spin correlation functions \label{ddss}}
\subsubsection{The density-density correlation function}
The density-density correlation function is defined to be
\beq \Pi_{jj'}(\tau) = \langle\widetilde{n}_j(\tau)\widetilde{n}_{j'}\rangle= \frac{Tr(e^{-\beta H} \widetilde{n}_j(\tau)\widetilde{n}_{j'})}{Z},\label{Pi1} \eeq
where
\barray 
\widetilde{n}_j(\tau) \equiv n_j(\tau)-\langle n_j \rangle;\;\;\;\;n_j \equiv \sum_\sigma X_j^{\sigma\sigma}. 
\earray
Expanding the exponentials in the density matrix and the time dependence of the number operator, we obtain
\beq \resizebox{1.15\hsize}{!}{$(\Pi_{jj'}(\tau)+n^2)(\frac{Z}{Z_0}) = \sum\limits_{a=0b=0}^\infty \frac{(\beta-\tau)^a}{a!} \frac{\tau^b}{b!} \sum\limits_{\substack{j_1j_1'\ldots j_nj_n'\\\sigma_1\ldots \sigma_n}} t_{j_1j_1'}\ldots t_{j_nj_n'}\langle X_{j_1'}^{\sigma_10}X_{j_1}^{0\sigma_1}\ldots X_{j_a'}^{\sigma_a0}X_{j_a}^{0\sigma_a}n_jX_{j_{a+1}'}^{\sigma_{a+1}0}X_{j_{a+1}}^{0\sigma_{a+1}}\ldots X_{j_n'}^{\sigma_n0}X_{j_n}^{0\sigma_n}n_{j'}\rangle_0$}, \label{Pi2} \eeq
where $n=a+b$. We shall now state the rules for calculating the $n^{th}$ order contribution to $\Pi_{jj'}(\tau)$. The proof of these rules runs along the same lines as the ones given for the thermodynamic potential and the Green's function. We will not give the full proof, but will merely point out a few key points particular to this case. First, we give the rules for drawing the diagrams $D_\Pi$, and evaluating their contributions $\Pi_{D_\Pi}$. The rules for $\Pi_{jj'}(\tau)$ will be given in terms of these diagrams and the partition function diagrams $D_i$. Rules for drawing and evaluating the $n^{th}$ order diagram $D_\Pi$:
\begin{itemize}
\item[1] If $n>0$, draw the diagram $D_\Pi$ in the same way as you would a connected diagram for $\frac{Z}{Z_0}$. Mark a site on this diagram to distinguish it from the other sites. If $n=0$, the only diagram is a single full site whose contribution is $\rho$. The single full site is then the site marked.
\item[2] Insert a factor of $\frac{\rho}{m}$ for each filled circle, and $1-\rho$ for each empty circle. Insert a factor of $t^n$.
\item[3] Define the time $\tau_{a.5}$ for $0\leq a \leq n$ to lie between $\tau_a$ and $\tau_{a+1}$ (recall also that higher numbers correspond to ``earlier" times when drawing the diagram). Then, insert a factor of $\sum\limits_{a=0}^n \frac{(\beta-\tau)^a}{a!} \frac{\tau^b}{b!} \ f(\tau_{a.5})$, where $b=n-a$, $f(\tau_{a.5})=1$ if the site marked in rule 1 is full {\bf at time} $\tau_{a.5}$, and $f(\tau_{a.5})=0$ otherwise. The site being full at a certain time means that either the last line on this site before this time entered the site, or it is a filled vertex whose earliest line occurs after this time.
\item[4] The sign and spin sum of the diagram are determined in the same way as for $\frac{Z}{Z_0}$.
\end{itemize}
The $n^{th}$ order contribution to $\Pi_{jj'}(\tau)$ is split into 2 pieces, $\Pi^{(n)}_{jj'}(\tau)=\Pi^{(n)}_{jj',a}(\tau)+\Pi^{(n)}_{jj',b}(\tau)$. Rules for calculating $\Pi^{(n)}_{jj',a}(\tau)$:
\begin{itemize}
\item[1] Choose one diagram $D_\Pi$, and $p \ge 0$ connected diagrams (not necessarily distinct) from $\frac{Z}{Z_0}$. The orders of the $p+1$ diagrams must add up to $n$. The $p+1$ diagrams are now components in this diagram.
\item[2] Multiply the contribution $\Pi_{D_\Pi}$ from $D_\Pi$ with the contributions $z_{nLD_1}$ \ldots $z_{nLD_p}$ from the $p$ components $D_1 \ldots D_p$.
\item[3] Fix any full site in the diagram $D_\Pi$ to be $j'$ on the lattice. Fix the marked site in the diagram $D_\Pi$ (from rule 1 of the rules for $D_\Pi$) to be $j$ on the lattice. Create all distinct $w^*$ from the components by forming $i_1(w^*)$ type 1 parentheses, \ldots , $i_{m_{w^*}}(w^*)=1$ type $m_{w^*}$ parentheses. There is only one outer parenthesis in each $w^*$ which encloses all $p+1$ components. The contribution of each $w^*$ is $\frac{(-1)^{i_1(w^*)+\ldots+i_{m_{w^*}}(w^*)}LS[w^*]}{\gamma(w^*)}$, where $\gamma(w^*)$ is the symmetry factor of $w^*$, and $LS[w^*]$ is the lattice sum of the overlapping of components represented by $w^*$. Note that in calculating $\gamma(w^*)$, $D_\Pi$ {\bf should not} be considered identical to any of the $D_1 \ldots D_p$. Sum the contributions over all $w^*$.
\item[4] Multiply the factor from 3 with the factor from 2.
\end{itemize}
Rules for calculating $\Pi^{(n)}_{jj',b}(\tau)$:
\begin{itemize}
\item[1] Choose one diagram $D_\Pi$, one connected diagram from $\frac{Z}{Z_0}$ (or the single full site with contribution $\rho$) denoted by $\widetilde{D}$, and $p \ge 0$ connected diagrams (not necessarily distinct) from $\frac{Z}{Z_0}$. The orders of the $p+2$ diagrams must add up to $n$. The $p+2$ diagrams are now components in this diagram.
\item[2] Multiply the contribution $\Pi_{D_\Pi}$ from $D_\Pi$ with the contribution $z_{nL\widetilde{D}}$ from $\widetilde{D}$, and the contributions $z_{nLD_1}$ \ldots $z_{nLD_p}$ from the $p$ components $D_1 \ldots D_p$.
\item[3] Fix any full site in the diagram $\widetilde{D}$ to be $j'$ on the lattice. Fix the marked site in the diagram $D_\Pi$ to be $j$ on the lattice. Create all distinct $w^*$ from the components by forming $i_1(w^*)$ type 1 parentheses, \ldots , $i_{m_{w^*}}(w^*)=1$ type $m_{w^*}$ parentheses. There is only one outer parenthesis in each $w^*$ which encloses all $p+2$ components. The contribution of each $w^*$ is $\frac{(-1)^{i_1(w^*)+\ldots+i_{m_{w^*}}(w^*)}LS[w^*]}{\gamma(w^*)}$, where $\gamma(w^*)$ is the symmetry factor of $w^*$, and $LS[w^*]$ is the lattice sum of the overlapping of components represented by $w^*$. Note that in calculating $\gamma(w^*)$, {\bf neither} $D_\Pi$ {\bf nor} $\widetilde{D}$ should be considered identical to any of the $D_1 \ldots D_p$. Sum the contributions over all $w^*$.
\item[4] Multiply the factor from 3 with the factor from 2.
\end{itemize}
The following observations went into deriving these rules. 
\begin{itemize}
\item[a] Number operators commute with all other operators (on different sites), and accommodate all spin species with equal coefficient. Therefore, they don't affect the sign or spin sum of a diagram. 
\item[b] Since the number operators don't create or destroy particles, $n_j$ and $n_{j'}$ don't have to occur in the same connected component. When they do appear in the same connected component, this is the component $D_\Pi$. When they appear in different connected components, $n_j$ appears in $D_\Pi$ while $n_{j'}$ appears in $\widetilde{D}$. 
\item[c] For the density-density correlation function, $D_\Pi$ plays the same role as $D_G$ plays in the Green's function. For $\Pi_{jj',a}(\tau)$, the combinatorial factors involved in distributing lines work out exactly as they do in \disp{etaGa}. For $\Pi_{jj',b}(\tau)$, the presence of $\widetilde{D}$ does not complicate matters since the operator $n_{j'}$ occurs to the right of all other operators in the expectation value in \disp{Pi2}. Hence, any combination of lines in the disconnected diagram can go into making $\widetilde{D}$, as is the case for the diagrams $D_1 \ldots D_p$ (but in contrast to $D_\Pi$). Therefore, $\widetilde{D}$ of order $\widetilde{n}$ gets the usual factor of $\frac{\beta^{\widetilde{n}}}{\widetilde{n}!}$. 
\item[d] Below \disp{z4}, we explain that the different components originally present in the diagram go into making the various ``generalized components" $w$. In the case when both $D_\Pi$ and $\widetilde{D}$ are present, they can either both go into making the same $w$ or go into different $w$'s. The former is accounted for by $\Pi_{jj',b}(\tau)$, while the latter is cancelled by $ (-)n^2$.
\end{itemize}  
\subsubsection{The spin-spin correlation function}
The spin-spin correlation function is defined to be
\beq \Pi^s_{jj'}(\tau) = \langle\widetilde{s^z_j}(\tau)\widetilde{s^z_{j'}}\rangle, \label{Pisz1} \eeq
where
\barray 
\widetilde{s^z_j}(\tau) \equiv s^z_j(\tau)-\langle s^z_j \rangle;\;\;\;\; s^z_j \equiv \sum_\sigma s(\sigma) X_j^{\sigma\sigma}. 
\earray
By $s(\sigma)$ we mean the spin corresponding to $\sigma$. For example, for spin-half particles, $s(\pm1)=\pm\frac{1}{2}$.
We note that for spin $l$ particles ($m=2l+1$):
\barray
\sum_\sigma s(\sigma) = 0;\;\;\;\;\; \sum_{\sigma} s^2(\sigma) = \frac{1}{3}l(l+1)m.
\earray
The first of these two equations tells us that $s^z_j$ and $s^z_{j'}$ must occur in the same ``loop" of the diagram. Therefore, the rules for the spin-spin correlation function can be obtained from those for the density-density correlation function by making the following simple changes.
\begin{itemize}
\item[1] $\Pi^s_{jj',b}(\tau) = 0$
\item[2] In the calculation of $D_{\Pi^s}$, mark any site $j$ and a full site $j'$ (see rule 1 for $D_\Pi$ for comparison). This is in contrast to $D_\Pi$ in which the choice of $j'$ did not affect the evaluation of $D_\Pi$ (or occurred in $\widetilde{D}$ rather than $D_\Pi$), and hence only came when calculating the lattice sum in rule 3 for $\Pi_{jj',a}(\tau)$ or $\Pi_{jj',b}(\tau)$.
\item[3] Rule 3 for the contribution of $D_{\Pi^s}$ is modified from that of  $D_\Pi$, so that there are now additional requirements for $f(\tau_{a.5})=1$. In the case that the last line on the site $j$ before the time $\tau_{a.5}$ enters $j$, this line must be in the same loop as the earliest line on the site marked $j'$. In the case that the site $j$ is full and its earliest line occurs after the time $\tau_{a.5}$, then this line must be in the same loop as the earliest line on the site marked $j'$. Otherwise, $f(\tau_{a.5})=0$. 
\item[4] Insert a factor of $\frac{1}{3}l(l+1)=\frac{1}{12}(m^2-1)$ into the contribution of the diagram $D_{\Pi^s}$.
\end{itemize}
We note that the above rules imply that there will only be even order contributions to the density-density and spin-spin correlation functions, since this is the case for the partition function. This was expected, since the number operator and spin operator conserve particle number.

\section{Conclusion}

In conclusion, we have developed a high-temperature series for the thermodynamic potential, the Green's function, and the time-dependent density-density and spin-spin correlation functions in the infinite-U Hubbard model. The  $n^{th}$ order contribution in $\beta t$ is given in terms of diagrams consisting of $n$ lines connecting vertices that can either be empty or full, corresponding to unoccupied and occupied sites on the lattice. The signature and spin sum of the diagram are evaluated with the help of a simple rule that increases the efficiency of computation, and enables results to be obtained for any number of spin species with no additional difficulty. The contribution of a diagram factors into a temporal part and a spatial part. The computation proceeds in two stages. In the first stage, the temporal part is evaluated for all of the connected diagrams. In the second stage, an arbitrary number of connected diagrams are combined into a ``generalized connected diagram". Its temporal part is the product of the temporal parts of its constituent connected diagrams, while its spatial part is the lattice sum corresponding to overlapping its constituent connected diagrams on the lattice. The linked cluster theorem is proved, enabling one to express the thermodynamic potential and the dynamical correlation functions as a sum over the generalized connected diagrams. 

This is an especially efficient way of doing the calculation because the temporal part of each constituent connected diagram, which is by far the most time-consuming part of the calculation, is evaluated only once. The rest of the complexity is taken care of by calculating lattice sums of overlappings of constituent connected diagrams. This should be contrasted with the Metzner approach \cite{Metzner}, in which a ``generalized connected diagram" (referred to in \cite{Metzner} as a connected diagram) is broken into constituent connected diagrams through the use of cumulants. The temporal contributions of the constituent connected diagrams are then evaluated and multiplied together {\bf every time} a generalized connected diagram is broken up. Therefore, the temporal contribution of a given constituent connected diagram is evaluated many times. What is gained by this is extreme simplicity in evaluating lattice sums. However, although complex, the lattice sum part of our calculation takes very little computation time, and even in high order calculations, can be done for a few minutes on a computer \cite{Ehsannew}. This is the essential reason why our method constitutes an improvement over those employed previously for the Green's function, and especially for the time-dependent density-density and spin-spin correlation functions.
We have used our method to calculate the Green's function to fourth order in $\beta t$ valid for $m$ spin species on a d-dimensional hypercube. Taking the $m\to \infty$ limit, we obtained expressions for the Dyson-Mori self-energy to fourth order in $\beta t$, valid for the case of an infinite number of spin species. This may have interesting connections to slave Boson techniques \cite{ReadNewns,Coleman,KotliarLiu} used for the study of this model.

\section{Acknowledgements}
I have greatly benefited from many stimulating and useful discussions with Sriram Shastry, which prompted me to work on this expansion, and which were especially helpful in the initial phases of this work. I would like to thank Ehsan Khatami for helpful comments. I would like to thank Sriram Shastry and Ehsan Khatami for a careful reading of the manuscript, and helpful suggestions. This work was supported by DOE under grant no. FG02-06ER46319.


\begin{thebibliography}{235}
\bibitem{ECFL} B. S. Shastry, Phys. Rev. Letts. {\bf107}, 056403 (2011).
\bibitem{Monster} B. S. Shastry, Phys. Rev. {\bf B 87} 125124 (2013).
\bibitem{ECFLDMFT} R. Zitko, D. Hansen, E. Perepelitsky, J.Mravlje, A. Georges, and B. S. Shastry,  arXiv:1309.5284  (2013).
\bibitem{larged} E. Perepelitsky and B. S. Shastry,  	arXiv:1309.5373, to appear Annals of Physics (2013).
\bibitem{Xengetal}  X. Deng, J. Mravlje, R. Zitko, M. Ferrero, G. Kotliar, and A. Georges, Phys. Rev. Letts. {\bf110}, 086401 (2013).
\bibitem{Nozieres} P. Nozi\`{e}res,
Theory of Interacting Fermi Systems (W. A. Benjamin, Amsterdam, 1964).
\bibitem{Domb} Phase Transitions and Critical Phenomena, edited by C. Domb and M. S. Green (Academic, London, 1974), Vol. 3.
\bibitem{Betts} Exact high temperature series expansions for the XY model, Canadian Journal of Physics, {\bf48}, 1566 (1970).
\bibitem{Plischke} M. Plischke, J. Stat. Phys. {\bf11}, 159 (1974).
\bibitem{KuboTada} K. Kubo and M. Tada, Progr. Theor. Phys. {\bf 69}, 1345 (1983); {\bf 71}, 479 (1984).
\bibitem{Metzner} W. Metzner, Phys. Rev. {\bf B 43}, 8549 (1991).
\bibitem{ECQL}  B. S. Shastry, Phys. Rev. {\bf B 81}, 045121 (2010).
\bibitem{Anatomy} B. S. Shastry, Phys. Rev. {\bf B 84}, 165112 (2011).
\bibitem{Asymm} B. S. Shastry,  Phys. Rev. Letts. {\bf 109}, 067004 (2012).
\bibitem{ECFLAM} B. S. Shastry , E. Perepelitsky, and A.C. Hewson, arXiv: 1307.3492 (2013).
\bibitem{Gweon} G.-H. Gweon, B. S. Shastry, and G. D. Gu, Phys. Rev. Lett. {\bf107}, 056404 (2011).
\bibitem{Kazue} K.Matsuyama and G.-H. Gweon, arXiv:1212.0299 (2013).
\bibitem{Moments} E. Khatami, D. Hansen, E. Perepelitsky, M. Rigol, and B. S. Shastry, Phys. Rev. {\bf B 87}, 161120 (2013).
\bibitem{DMFThighT} Lorenzo De Leo, Jean-SŽbastien Bernier, Corinna Kollath, Antoine Georges, and Vito W. Scarola, Phys. Rev. {\bf A 83}, 023606 (2011).
\bibitem{Pairault} Stephane Pairault, David Senechal, A.-M. S. Tremblay, Eur. Phys. J. {\bf B 16}, 85 (2000).
\bibitem{ReadNewns} N. Read, D.M. Newns, J. Phys. {\bf C 16}, 3273 (1983). 
\bibitem{Coleman} P. Coleman, Phys. Rev. {\bf B 28}, 5255 (1983).
\bibitem{KotliarLiu} G. Kotliar and J. Liu, Phys. Rev. {\bf B 38}, 5142 (1988).
\bibitem{EhsanMetzner} ``Linked-Cluster Expansion of the GreenÕs function of the infinite-U Hubbard Model", E. Khatami, E. Perepelitsky, M. Rigol, and B. S. Shastry, to be published (2013).
 \bibitem{Ehsannew} ``A study of the phase transitions of the infinite-U Hubbard Model", E. Khatami, E. Perepelitsky, M. Rigol, and B. S. Shastry, to be published (2013).
 \bibitem{jaxodraw} Diagrams in the paper created through the use of ``JaxoDraw", D. Binosi,
L. Theu{\ss}l, Computer Physics Communications, {\bf Volume 161}, Issues 1-2, 1 August 2004, Pages 76-86
 



\end{thebibliography}
\end{document}